\documentclass[preprint2]{emulateapj}

\newcommand{\chn}{{\it Chandra}}
\newcommand{\asec}{\arcsec}

\usepackage{latexsym}
\usepackage{graphicx}
\usepackage{longtable}
\usepackage{array}
\usepackage{natbib}
\usepackage{morefloats}
\usepackage{verbatim}
\usepackage[colorlinks=true, pdfstartview=FitV, linkcolor=blue,citecolor=blue, urlcolor=blue]{hyperref}
\usepackage[bf]{subfigure}
\usepackage{multirow}

\shorttitle{The 3CR \chn\ extragalactic survey at 1.0$<z<$1.5}
\shortauthors{C. Stuardi et al. 2018}

\begin{document}

\title{The 3CR \chn\ snapshot survey: extragalactic radio sources \\ with redshifts between 1 and 1.5}

\author{
C. Stuardi\altaffilmark{1,2,3,4},
V. Missaglia\altaffilmark{5},
F. Massaro\altaffilmark{2,3,6}, 
F. Ricci\altaffilmark{7,8},
E. Liuzzo\altaffilmark{4},
A. Paggi\altaffilmark{8}, \\
R. P. Kraft\altaffilmark{8},
G. R. Tremblay\altaffilmark{8},
S. A. Baum\altaffilmark{9,10},
C. P. O'Dea\altaffilmark{9,11}, \\
B. J. Wilkes\altaffilmark{8},
J. Kuraszkiewicz\altaffilmark{8},
W. R. Forman\altaffilmark{8}
\&
D. E. Harris\altaffilmark{8,+}
}

\altaffiltext{1}{Dipartimento di Fisica e Astronomia, Universit\`a di Bologna, via Piero Gobetti 93/2, 40129 Bologna, Italy}
\altaffiltext{2}{Dipartimento di Fisica, Universit\`a degli Studi di Torino, via Pietro Giuria 1, 10125 Torino, Italy.}
\altaffiltext{3}{INAF-Osservatorio Astrofisico di Torino, via Osservatorio 20, 10025 Pino Torinese, Italy}
\altaffiltext{4}{Istituto di Radioastronomia, INAF, via Gobetti 101, 40129, Bologna, Italy.}
\altaffiltext{5}{Department of Physical Sciences, University of Napoli Federico II, via Cinthia 9, 80126 Napoli, Italy.}
\altaffiltext{6}{Istituto Nazionale di Fisica Nucleare, Sezione di Torino, 10125 Torino, Italy.}
\altaffiltext{7}{Dipartimento di Matematica e Fisica, Universit\`a Roma Tre, via della Vasca Navale 84, I-00146, Roma, Italy.}
\altaffiltext{8}{Smithsonian Astrophysical Observatory, 60 Garden Street, Cambridge, MA 02138, USA.}
\altaffiltext{9}{University of Manitoba,  Dept. of Physics and Astronomy, Winnipeg, MB R3T 2N2, Canada.}
\altaffiltext{10}{Center for Imaging Science, Rochester Institute of Technology, 84 Lomb Memorial Dr., Rochester, NY 14623, USA.}
\altaffiltext{11}{School of Physics \& Astronomy, Rochester Institute of Technology, 84 Lomb Memorial Dr., Rochester, NY 14623, USA.}
\altaffiltext{+}{Dan Harris passed away on December 6th, 2015. His career spanned much of the history of radio and X-ray astronomy. His passion, insight, and contributions will always be remembered. A significant fraction of this work is one of his last efforts.}

\begin{abstract} 

\indent The aim of this paper is to present an analysis of newly acquired X-ray observations of 16 extragalactic radio
sources, listed in the Third Cambridge Revised (3CR) catalog, and not previously observed by \chn . Observations were performed 
during \chn\ Cycle 17, extending X-ray coverage for the 3CR extragalactic catalog up to $z$=1.5. Among the 16 targets, two lie at $z<$0.5 (i.e., 3CR\,27, at $z$=0.184 and 3CR\,69, at $z$=0.458), all the remaining 14 have redshifts 
between 1.0 and 1.5. In the current sample there are three compact steep spectrum (CSS) sources, three quasars and an FR\,I radio galaxy, while the other nine are FR\,II radio galaxies. All radio sources have an X-ray counterpart. We measured nuclear X-ray fluxes as well as X-ray emission associated with radio jet
knots, hotspots or lobes in three energy bands: soft (0.5-1 keV), medium (1-2 keV) and hard (2-7 keV). We also performed standard X-ray spectral analysis for the four brightest
nuclei. We discovered X-ray emission associated with: the radio lobe of 3CR\,124; a hotspot of the quasar 3CR\,220.2; another hotspot of the radio galaxy 3CR\,238; and the jet knot of 3CR\,297. We also detected extended X-ray emission around the nuclear region of 3CR\,124 and 3CR\,297 on scales of several tens of kpc. Finally, we present an update on the X-ray observations performed with \chn\ and {\it XMM-Newton} on the entire 3CR extragalactic catalog.

\end{abstract}

\keywords{galaxies: active --- X-rays: general --- radio continuum: galaxies}

\section{Introduction}
\label{sec:intro}

The last revised version of the Third Cambridge catalog (3CR) of extragalactic radio sources \citep{spinrad85} remains one of the best studied samples of radio-loud active galactic nuclei (AGNs) being critical for statistical analysis of their properties \citep[see e.g.,][]{schields99,tadhunter16}. The 3CR catalog is a 178 MHz radio flux-limited sample, with a 9 Jy cut-off, that covers the redshift range 0.0005$<z<$2.5. As a low frequency selected sample, it is unbiased with respect to viewing angle, X-ray properties (e.g., total X-ray nuclear flux and spectrum), and optical spectroscopic classification of the sources.

In recent years, the large majority of the 298 3CR extragalactic sources have been observed with several photometric and spectroscopic surveys, in the radio, infrared (IR) and optical wavebands, so that a vast suite of data is now available to facilitate multifrequency analyses. Radio images with arcsecond resolution for the majority of the 3CR sources are available from the National Radio Astronomy Observatory Very Large Array (VLA) Archive Survey (NVAS) and in the archive of the MERLIN observatory. Recently, several snapshot surveys of 3CR radio sources were carried out using the Spitzer satellite \citep[see e.g.,][]{ramirez14, dicken14, ghaffari17} and the Hubble Space Telescope (HST) \citep[see e.g.,][]{privon08,tremblay09}, reaching a coverage of about  90\% of the entire extragalactic catalog including new observations of high redshift sources at $z>$1.0 \citep[see also][]{leipski10, hilbert16}. Several high redshift 3CR radio galaxies and quasars were detected in the far-infrared also by the Herschel Space Observatory \citep{podigachoski16}. Moreover, ground-based spectroscopic observations were carried out with the Telescopio Nazionale Galileo \citep{buttiglione09,buttiglione11}.

To extend the wavelength coverage of the 3CR catalog, we started an X-ray snapshot survey  with \chn , the only X-ray facility with angular resolution comparable to that at optical and radio frequencies \citep[see][for a recent review]{massaro15}. The main aims of the \chn\ snapshot survey can be summarised as: {\it i)} studying the X-ray emission, as a function of $z$ and radio power, arising from jet knots, hotspots and nuclei of radio sources, {\it ii)} investigating the nature of their large-scale environment and {\it iii)} searching for observational evidence of AGN interactions with the hot gas in galaxies, groups and clusters of galaxies. This interaction, known as {\it feedback}, is observed, for example, in cool-core galaxy clusters where extended structures (e.g., lobes) of radio sources, expand into the X-ray emitting intracluster medium creating cavities filled with radio-emitting plasma \citep[see e.g.,][]{fabian12, kraft12}. \chn\ observations also allow us to search for new galaxy clusters via the presence of extended X-ray emission unrelated to the radio structures on kpc and Mpc scales around the sources \citep[see e.g.,][]{belsole07,ineson13,mannering13}.

Before Cycle 9, only $\sim60\%$ of 3CR extragalactic sources were observed by \chn : the snapshot survey allowed us to observe 113 more targets and all those with $z<$1.5 have at least a snapshot ($<$20 ks exposure) observation available in the \chn\ archive\footnote{http://cda.harvard.edu/chaser/} to date. In the previous subsamples, observed during Cycles 9, 12, 13 and 15, we detected significant X-ray emission from all but two active nuclei and from 22 hotspots \citep[see][]{massaro10, massaro12, massaro13, massaro17}. Extended X-ray emission, possibly arising from the hot gas of a surrounding galaxy group or cluster, was also detected around 16 sources. Moreover, the \chn\ snapshot survey inspired follow-up observations on several interesting objects such as 3CR\,171 \citep{hardcastle10}, 3CR\,305 \citep{hardcastle12} and the tailed radio galaxy 3CR\,89 \citep{dasadia16}. Our final goal of completing the \chn\ observational coverage of the entire 3CR extragalactic catalog could be achieved in the next Cycle, with the observation of the remaining 9 sources. 

Here, we present the X-ray analysis of \chn\ observations carried out during Cycle 17, including 16 3CR radio sources previously unobserved by \chn . Among the 16 targets, 14 are in the redshift range 1.0-1.5, while revised spectroscopic observations yielded an updated redshift estimate for the other two: 3CR\,27, $z$=0.184, and 3CR\,69, $z$=0.458 \citep{hiltner91}. 

The paper is organised as follows: a brief description of the observations and data reduction procedures is given in \S~\ref{sec:obs} while general results and source details are described in \S~\ref{sec:results}. A brief summary and conclusions are given in \S~\ref{sec:summary}. Finally, we present in Appendix~\ref{sec:appendixA} X-ray images with radio contours overlaid for all the sources in the current high redshift sample, and in Appendix~\ref{sec:appendixB} the updated summary of X-ray observations, including \chn\ and {\it XMM-Newton} informations, for the entire 3CR extragalactic portion of the catalog.

Throughout, we assumed a flat cosmology with $H_0=69.6$ km s$^{-1}$ Mpc$^{-1}$,
$\Omega_{M}=0.286$ and $\Omega_{\Lambda}=0.714$ \citep{bennett14}, and we adopted cgs units, unless stated otherwise. Spectral indices, $\alpha$, are defined by the flux density, S$_\nu \propto \nu^{-\alpha}$.

\begingroup
\setlength{\tabcolsep}{3pt}
\begin{table*} 
\footnotesize 
\caption{Source List of the \chn\ Cycle 17 Snapshot Survey of  3CR Radio Sources}
\label{tab:main}
\begin{center}
\begin{tabular}{lccclccclrcc}
\hline
\hline
3CR & Class\tablenotemark{a} & R.A. (J2000)\tablenotemark{b} & Dec. (J2000)\tablenotemark{b} & $z$\tablenotemark{c} & D$_L$ & Scale & N$_{H,Gal}$\tablenotemark{d} & m$_v$\tablenotemark{e} & S$_{178}$\tablenotemark{f}  & \chn\  & Obs. Date \\ & & (hh mm ss) & (dd mm ss) & & (Mpc) & (kpc/\arcsec) & (10$^{20}$~cm$^{-2}$) & & (Jy) & ObsID & (yyyy-mm-dd) \\ 
\hline 
\noalign{\smallskip}
27*   &  FR\,II-HERG & 00 56 01.0~~\, & +68 22 30~~~\, & 0.184  & 900 & 3.113  & 41.2 & 18.2 & 26.5 & 18090  & 2016-03-15 \\ 
69   &  FR\,II-HERG & 02 38 02.673 & +59 11 50.56 & 0.458  & 2577 & 5.877  & 68.5 & 19.0 & 20.9 & 18092  & 2016-04-09 \\
\hline 
\noalign{\smallskip}
36   &  FR\,II-HERG & 01 17 59.555 & +45 36 22.43 & 1.301  & 9293 & 8.510  & 6.35 & 20.0$^+$ & 8.2 & 18091  & 2016-03-05 \\
119     &  CSS & 04 32 36.505 & +41 38 28.44 & 1.023 & 6893 &  8.166  & 36.7 & 20.0$^+$ & 15.7 & 18093  & 2015-12-20 \\                 
124     &  FR\,II-HERG & 04 41 59.129 & +01 21 01.10 & 1.083 & 7399 & 8.267 & 8.75 & 22.0 & 10.3 & 18094  & 2015-12-15 \\ 
173     &  CSS\,-\,HERG & 07 02 20.474 & +37 57 22.54 & 1.035 & 6994 & 8.187  & 9.42 & 21.3$^+$ & 8.7 & 18095  & 2015-12-03 \\ 
194     &  FR\,II-HERG  & 08 10 03.636 & +42 28 05.02 & 1.184   & 8266 & 8.402  & 4.76 & 20.0$^+$ & 9.9  & 18096  & 2015-12-17 \\
208.1  &  QSO & 08 54 39.373 & +14 05 52.06 & 1.020$^+$  & 6868  & 8.160 & 3.35 & 20.0$^+$ & 8.1 & 18097  & 2016-03-01 \\ 
220.2  &  QSO & 09 30 33.541 & +36 01 25.14  & 1.157$^+$ & 8033 &  8.370  & 1.20 & 19.0$^+$ & 7.2 & 18098  & 2016-02-08 \\ 
222     &  FR\,I & 09 36 32.023 & +04 22 10.18 & 1.339  & 9632 & 8.535 & 3.10 & 23.0$^+$ & 11.3 & 18099  & 2016-01-15 \\
230*     &  FR\,II-HERG & 09 51 58.8~~\, & -- 00 01 27~~~\, & 1.487   & 10970 & 8.599 & 3.63 & \,\, -- & 21.1 & 18100  & 2016-01-13 \\ 
238     &  FR\,II-HERG & 10 11 00.350 & +06 24 39.02 & 1.405$^+$  & 10225 & 8.570       & 1.96 & 22.5$^+$ & 16.6 & 18101  & 2016-01-28 \\
255     &  FR\,II(?)-HERG & 11 19 25.292 &    -- 03 02 51.36 & 1.355  & 9775 & 8.545  & 4.29 & 23.0 & 12.5 & 18102  & 2016-02-07 \\
297*     &  QSO & 14 17 24.0~~\, & -- 04 00 48~~~\, & 1.406$^+$  & 10234 & 8.571 & 3.29 & 21.9$^+$ & 10.3 & 18103  & 2016-03-27 \\ 
300.1  &  FR\,II-HERG  & 14 28 31.230 &    -- 01 24 07.04 & 1.159  & 8050 & 8.372       & 3.13 & 19.0$^+$ & 14.1 & 18104  & 2015-12-21 \\
305.1*  &  CSS\,-\,LERG & 14 47 09.5~~\, & +76 56 21~~~\, & 1.132$^+$  & 7817 & 8.338 & 3.12 & 21.4$^+$ & 4.6 & 18105  & 2016-08-07 \\
\noalign{\smallskip}
\hline
\end{tabular}\\

\end{center}
\textbf{Notes.} The two sources above the line are outside the redshift range 1.0$<z<$1.5.\\
\tablenotemark{a} The ``Class'' column contains both a radio morphology descriptor (Fanaroff-Riley class I or II), quasar (QSO) or Compact Steep Spectrum (CSS) and 
the optical spectroscopic designation, LERG, ``Low Excitation Radio Galaxy'', HERG, ``High Excitation Radio Galaxy'', when present in the literature.\\
\tablenotemark{b} The celestial positions listed are those of the radio cores which we used to register the X-ray images, except for the four sources lacking an obvious radio nucleus: 3CR\,27, 3CR\,230, 3CR\,297 and 3CR\,305.1, labelled with ``\,*\,'', for which the coordinates are taken from the NASA/IPAC Extragalactic Database (NED). \\
\tablenotemark{c} Redshift measurements are taken from Spinrad \citep{spinrad85} or from NED. Data from Spinrad are labeled with ``$^+$''. \\
\tablenotemark{d} Galactic neutral hydrogen column densities $N_{H,Gal}$ \citep{kalberla05}.\\
\tablenotemark{e} $m_v$ is the visual magnitude from Spinrad \citep{spinrad85} or from NED. Data from Spinrad are again labeled with ``$^+$''.  The value of visual magnitude reported by NED for 3CR\,230 is affected by a star nearby in projection on the sky and it is not reported here.\\
\tablenotemark{f} S$_{178}$ is the flux density at 178 MHz \citep{spinrad85}.\\
\end{table*} 
\endgroup

\section{Observations, Data Reduction, and Analysis}
\label{sec:obs}

We adopted the data reduction and analysis procedures described in previous works, thus only basic details are reported here  \citep[see e.g.][ for a complete description]{massaro09a,massaro11}. We compared radio images, ranging from 1.4 to 14.9 GHz, and \chn\ observations searching for a spatial coincidence between X-ray emission and extended radio structures (i.e., jet knots, hotspots or lobes).

\subsection{Radio Observations}
\label{sec:Radobs}

Radio images were retrieved from the NVAS\footnote{https://archive.nrao.edu/nvas} and the DRAGN\footnote{http://www.jb.man.ac.uk/atlas/} webpage.

For the peculiar case of 3CR\,297 (see \S~\ref{sec:results}), we also computed the radio spectral index map, using archival observations at 4.85 and 8.4 GHz. The radio spectral index, $\alpha_R$, is a diagnostic tool to obtain useful informations on the particle energy distribution of electrons emitting via synchrotron emission. Spatial variations in spectral index across the radio structure were used to distinguish different regions of 3CR\,297, since they could indicate differences in the underlying physical processes occurring in cores and in jets: radio emission arising from compact cores has typically a flat radio spectrum (i.e., $\alpha_R\lesssim0.5$), while that from extended structures is characterised by a steeper spectrum \citep[see e.g.,][]{pauliny68}.

Archival VLA data analysed are from project ID AE0059 (4.85 GHz) and AV0164 (8.4 GHz) 
and were obtained on 30 October 1988 and 11 May 1990, respectively. In both observing runs the VLA configuration was A. 
We performed post-correlation processing and imaging with the Common Astronomy Software Applications (CASA). 

We first calibrated the images in amplitude and phase with the primary calibrators 1127-145 (4.85 GHz) 
and 1328+307 (8.4 GHz). We set the uv-range for the computation of gain solutions in the latter case to a maximum of 400 k$\lambda$, following the specification on the VLA fringe calibrators\footnote{https://science.nrao.edu/facilities/vla/observing/callist}. For the 4.85 GHz data, we flagged the antenna VA20 in the RR correlation in the bin time 14:28:55.0 to remove the bad response and achieve better gain solutions. Then, a self-calibration procedure was iteratively applied to improve the signal-to-noise, using the CASA task \texttt{gaincal + applycal}. 
We obtained the model image through the \texttt{clean} task using the weighting parameter set on \texttt{natural} first, and \texttt{uvtaper} at 
the end, to model also the extended emission. During this procedure we flagged additional deviating data (at 8.4 GHz the antenna VA22 for time less than 06:42:00 and VA13 between 07:00:0 and 07:01:15). 

We computed the two final continuum maps with the same beam (0.98\asec, 0.88\asec, -35$^{\circ}$, task \texttt{clean} using \texttt{uvtaper}) and we re-gridded both to the same cell size (task \texttt{imregrid}) to properly compute the spectral index image for pixels with values $>3\sigma$ with the relation: $\alpha_R= - log(S_1/S_2)/log(\nu_1/\nu_2)$ with $\nu_1=4.85$ GHz, $\nu_2=8.4$ GHz, and $S_{1,2}$ being the respective flux at the two frequencies (task \texttt{immath}). The resulting average rms of 3CR 297 radio map was ~0.4 (~0.15) mJy/beam and the peak flux was 0.510 (0.299) Jy at 4.85 (8.4) GHz. Typical uncertainties for flux density measurements in 3CR 297 radio maps are $\sim$10\% for 4.85 GHz and $\sim$3\% for 8.4 GHz.

\subsection{X-ray Observations}
\label{sec:Xobs}

The 3CR source sample observed during Cycle 17 is listed in
Table~\ref{tab:main}, together with the basic parameters of each source (e.g., radio and optical classification, celestial coordinates, redshift, visual magnitude and radio flux). Sources were observed for a nominal exposure time of 12 ksec, but actual livetimes
are given in Table~\ref{tab:fluxes}. All \chn\ observations were performed with the ACIS-S back illuminated chip in VERY FAINT mode to provide high sensitivity and a low background level. The four chips turned on were I2, I3, S2, and S3 with the nominal aim point centered on S3.  

We performed the data reduction following the standard reduction
procedure described in the \chn\ Interactive Analysis of Observations
(CIAO) threads\footnote{http://cxc.harvard.edu/ciao/guides/index.html}, using
CIAO version 4.9 and the \chn\ Calibration Database (CALDB) version 4.7.3. Level 2 event 
files were generated using the \texttt{acis\_process\_events} task, and events were filtered 
for grades 0,2,3,4,6 and corrected for bad pixels. Light curves for every data set
were extracted and checked for high background intervals, but none were found.

Since the on-axis width of the point spread function of the \chn\ telescope is smaller than the 
size of the ACIS pixels (0\arcsec.492), to recover the native angular resolution, it is necessary to avoid
the under-sampling. This is achieved by re-gridding all the images 
to 1/4 of the native size to obtain a common pixel size of 0\arcsec.123.

When possible, we registered X-ray images, changing the
appropriate keywords in the header of the event file so as to align the nuclear
X-ray position with that of the radio. In most cases, the total angular shift (reported for each registered source in Table~\ref{tab:fluxes}) was less than 1\arcsec, as occurred in all previous sources analysed and collected in the 
XJET website\footnote{http://hea-www.harvard.edu/XJET/} \citep{massaro11}. The registration facilitates accurate searches for X-ray emission from within the extended radio structures
by comparing radio and X-ray data with similar angular resolution. The distribution of the offsets in Table~\ref{tab:fluxes} is consistent with the level of uncertainty in the Chandra absolute astrometry (0\asec .8 at 90\% confidence, 1\asec .4 at 99\%\footnote{http://cxc.harvard.edu/cal/ASPECT/celmon/}). We have also confirmed by eye the identification of the X-ray, radio and optical nuclei in all cases.

\begin{table*}
\caption{Nuclear X-ray Fluxes}
\label{tab:fluxes}
\begin{center}
\begin{tabular}{lccrcrrrrccr}
\hline
\hline
3CR  &  Shift\tablenotemark{a}   & LivTim\tablenotemark{b}&  $N_2$\tablenotemark{c}& Ext. Ratio\tablenotemark{d}& f(Soft)\tablenotemark{e} & f(Medium)\tablenotemark{e}  &  f(Hard)\tablenotemark{e}    & f(Total)\tablenotemark{e}        & HR\tablenotemark{f} & L$_X$\tablenotemark{g} \\ 
    &  (\arcsec)   & (ks)                                 & (counts)                             &                                                   & 0.5-1~keV & 1-2~keV   & 2-7~keV & 0.5-7~keV & & (10$^{44}$~erg~s$^{-1}$) \\
\hline 
\noalign{\smallskip}
27  & --- & 11.91   &  61(8)   &  0.90(0.06)  & ---  &  2.3(1.0)   & 86(11)  & 88(11)   &   $<$\,0.94  &    0.09(0.01)      \\    
69  & 0.62 & 11.92   &   497(22)     &  0.94(0.01)   & 9(2)        &  74(5)       & 350(20)  & 433(21)  & 0.65(0.03)    &   3.4(0.2)  \\
\hline 
\noalign{\smallskip}
36  &  0.66  & 11.91   &   335(18)     &  0.94(0.01)   & 27(4)  & 65(5) & 143(13)  & 235(15)  &  0.37(0.05) &   24.3(1.6)   \\
119  & 0.60 & 11.90   &   383(20)     &  0.97(0.01)   & 18(4) & 64(5) & 220(16) & 303(17) & 0.54(0.04)  &   17.2(0.9) \\
124  & 0.48  & 11.92   &  21(5)    &   0.49(0.09)  & 1.5(1.0) & 0.5(0.5) & 22(5) & 24(5) & $<$\,0.96 &  1.6(0.3)   \\
173  & 1.10 & 12.40   &  124(11)    &  0.96(0.03)   & 11(3) & 21(3) & 43(7) & 76(8) & 0.34(0.09) &  4.4(0.5)  \\
194 &  0.36  & 11.91   &    125(11)     &  0.87(0.03)   & 0.8(0.8) & 18(3) & 89(10) & 108(11) & 0.67(0.05) &  8.8(0.9)  \\
208.1 &  0.73  & 11.91   &   228(15)     & 0.92(0.02)    & 25(4) & 35(4) & 102(11) & 162(12) & 0.48(0.06) &   9.1(0.7)  \\
220.2 &  0.44  & 11.92   &   684(26)     &   0.94(0.01)   & 78(7) & 119(7) & 273(18) & 470(20) & 0.39(0.04) &  36.3(1.5) \\
222 &  0.68  & 11.91   &   3(2)      &  ---   & --- & 0.8(0.6) & 1.5(1.5 ) & 2.3(1.6)  & --- &   0.3(0.2)  \\
230 & --- & 12.35   &   25(5)   &  0.7(0.1)   &  --- & 2.4(1.0) & 21(5) & 24(5) & $<$\,0.80 &  3.5(0.7)   \\
238 & 0.13 & 11.91   &   23(5)    &  0.5(0.1)   & 2.0(1.0) & 3.5(1.2) & 8(3) & 14(3) & ---  &  1.8(0.4)  \\
255 &  0.73 & 11.92   &   5(2)     & ---   &  0.7(0.7) & 1.0(0.6) & 1.4(1.4) & 3.1(1.7) & --- &   0.4(0.2)  \\
297 &  0.54  & 12.34   &   17(4)      &  0.24(0.06)   & 1.3(0.9) & 2.5(0.9) & 8(3) & 12(3) & --- &  1.5(0.4)  \\
300.1 &  0.24  & 11.92   & 19(4)    &  0.42(0.09)   & 0.6(0.6) & 0.8(0.6) & 22(6) & 23(6) & $<$\,0.93  & 1.8(0.5)     \\   
305.1  & ---  & 11.91   &   16(4)     &  0.7(0.2)   & 0.8(0.8) & --- & 20(5) & 20(5) & \, $<$\,0.92* &  1.5(0.4)  \\
\noalign{\smallskip}
\hline
\end{tabular}\\

\end{center}
\textbf{Notes.} Fluxes are given in units of 10$^{-15}$~erg~cm$^{-2}$~s$^{-1}$. Values in parentheses are 1$\sigma$ uncertainties.\\
\tablenotemark{a}{ Angular shift imposed to the X-ray image to align the nuclear X-ray position with that of the radio. The symbol  ``\,--\,'' marks the sources without astrometric registration.} \\
\tablenotemark{b}{ LivTim is the observation live time.} \\
\tablenotemark{c}{ Total number of counts within a circle of radius r=2\asec\ centered on the source position. The uncertainties given in parentheses are computed as $\sqrt{total-number-of-counts}$. }\\
\tablenotemark{d}{ ``Extent Ratio'': the ratio of the net counts (i.e., background subtracted) in the r=2\asec\ circle to the  net counts in the
r\,=\,10\arcsec\ circle.  Values less than 0.5 indicate the presence of extended emission around the nuclear component of 3CR\,124, 3CR\,297, and 3CR\,300.1. We did not report values of ``Extent Ratio'' for 3CR\,222 and 3CR\,255, having less than 9 counts in the 2\asec\ circle . A 1$\sigma$ uncertainty is derived from the Poisson uncertainties on the number of counts in the circular region of radius 2\asec\ and in the annular region between the radii 2\asec\ and 10\asec , taking into account the covariance of the terms and the uncertainty on the number of background counts measured on the CCD.}\\
\tablenotemark{e}{ Fluxes extracted from the flux maps in the three energy bands (0.5-1 keV, 1-2 keV, 2-7 keV) and the total flux in the energy range 0.5-7 keV. The uncertainties are derived from the relative uncertainties on the number of counts in the source and background regions, added in quadrature.  The symbol  ``\,--\,'' indicates that no counts were detected in the photometric region in the respective energy range.}\\
\tablenotemark{f}{ The observed nuclear fluxes have been used to determine the hardness ratios $HR$ with the relation: $(H-M)/(H+M)$, where $H$ and $M$ are the X-ray fluxes in the hard and the medium bands, respectively. We computed the $HRs$ for those sources having more than 9 counts in the hard and medium band, while we only estimated the upper bound of the HR value for those sources with 9 or more counts in the hard band alone. The $HR$ upper bound for the source 3CR\,305.1, labelled with ``\,*\,'', was computed with the soft X-ray flux $S$ (i.e., $(H-S)/(H+S)$) due to the lack of counts detected in the medium band. The uncertainties have been derived from the X-ray flux uncertainties.}\\
\tablenotemark{g}{ X-ray luminosity in the range 0.5 to 7 keV, computed using the values of D$_L$ given in Table~\ref{tab:main}. A 1$\sigma$ uncertainty is derived from the uncertainties on the flux.}\\
\end{table*}

\subsubsection{Flux maps}
\label{sec:fluxmaps}

We created flux maps in three energy ranges: 0.5-1 keV
(soft band), 1-2 keV (medium band) and 2-7 keV (hard band), by
filtering the event file with the appropriate energy range and
taking into account exposure time and effective area. Since the effective area is a function of the photon energy, we created monochromatic exposure maps with nominal energies, $E_{nominal}$, of 0.8 keV, 1.4 keV, and 4 keV for the soft,
medium and hard band, respectively. To obtain flux maps in units of ergs~cm$^{-2}$~s$^{-1}$~pixel$^{-1}$, 
we multiplied each event by the nominal energy of its respective band. Throughout the paper, we used the re-gridded pixels of size 0\asec.123.

To measure observed fluxes for nuclei and for any X-ray detected component, we chose an appropriate circular region.  For nuclei, we used circular regions of 2\arcsec\ radius (see Table~\ref{tab:fluxes}), while the radius of the circular regions used for other detected components is reported in Table~\ref{tab:jets}. X-ray fluxes were computed using the flux maps for each source region in the three energy ranges previously defined. Since the nominal energy was used only to recover the correct units in the flux maps, for each particular region we applied a correction by multiplying the flux by a factor of $E_{average}/E_{nominal}$, where $E_{average}$ is the mean energy computed in the photometric aperture, for the soft, medium and hard band, respectively. This correction ranged from a few to $\sim$15\%. For each radio feature, we chose two adjacent source-free background regions, with the same size described above, to avoid contaminating X-ray emission (as well as radio emission) and to sample both sides of the extended radio structure. The one $\sigma$ uncertainties on the fluxes were computed using the relative uncertainties on the number of counts, estimated assuming a Poisson distribution and taking into account the background. The background level in the \chn\ images is sufficiently low that small changes in the background regions used does not result in significant changes in the derived flux uncertainties.

We used the nuclear fluxes measured with this procedure to determine the X-ray hardness ratios ($HRs$) through the simple relation: $(H-M)/(H+M)$, where $H$ and $M$ are the X-ray fluxes in the hard and the medium bands, respectively. In Table~\ref{tab:fluxes}, the uncertainties on the observed value of $HR$ were propagated from the uncertainties of $H$ and $M$, taking into account the covariance terms \citep[see e.g.,][]{park06}. We did not use the soft band flux values because they are the most affected by Galactic absorption, except in the case of 3CR\,305.1 where no counts were detected in the medium band (see Table~\ref{tab:fluxes}). We did not compute the $HRs$ for those sources having less than 9 counts in each band, while for those with high counts in the hard band alone we estimated the upper bound for the $HR$ values.

Finally, we computed X-ray luminosities using the values of D$_L$ given in Table~\ref{tab:main}. X-ray fluxes and luminosities for the nuclei are given in Table~\ref{tab:fluxes} while those of the detected components are given in Table~\ref{tab:jets}, together with their detection significance evaluated adopting Poisson statistics. In the latter table, the component classification is indicated by a letter (i.e., hotspot $h$, lobe $l$, jet knot $k$, respectively), a cardinal direction as viewed from the nucleus and a number indicating the distance from the nucleus in arcseconds \citep[as in][]{massaro11}.

\begin{table*} 
\caption{Radio components with X-ray Detections}
\label{tab:jets}
\begin{center}
\begin{tabular}{llcccrrrrc}
\hline
\hline
3CR                     &Component\tablenotemark{a}& Radius\tablenotemark{b}& Total\,(bkg)\tablenotemark{c}& Detection &   
f(Soft)\tablenotemark{e} & f(Medium)\tablenotemark{e}  &  f(Hard)\tablenotemark{e}    & f(Total)\tablenotemark{e}    &L$_X$\tablenotemark{f}\\ 
                          &                                               & (arcsec)                         &           (counts)                                      & Significance\tablenotemark{d} &
0.5-1~keV         &       1-2~keV        &     2-7~keV      &        0.5-7~keV                             &10$^{44}$~erg~s$^{-1}$\\
\hline 
\noalign{\smallskip}
124        & $l\;-w\, 8$ &  2.5  &   5(1)* & $>$3$\sigma$  & --- &  1.2(0.7) & 1.9(1.9)  &  3(2) &  0.2(0.1)\\
 220.2     & $h-s\, 5$ & 2.0 &  11(0.3) & $>$7$\sigma$	 &  1.8(1.1)  & 0.9(0.7) & 4(3) & 7(3) & 0.5(0.2)\\
 238       & $h-n\, 3$ & 2.0 & 9(0.2) & $>$7$\sigma$ 	 & 1.3(0.9)  & 1.2(0.6) & 2.9(1.7) & 5(2) & 0.7(0.3)\\
 297       & $k-w\,4$ & 2.0 & 33(0.2) & $>$7$\sigma$	 & 3.9(1.8)  & 5.6(1.5)  & 9(3) & 19(4) & 2.4(0.5)\\
 \noalign{\smallskip}
\hline
\end{tabular}\\

\end{center}
\textbf{Notes.} Fluxes are given in units of 10$^{-15}$~erg~cm$^{-2}$~s$^{-1}$.\\
\tablenotemark{a} The component designation is comprised
of a letter indicating the classification (i.e., hotspot $h$, lobe $l$, jet knot $k$), 
a cardinal direction (as viewed from the nucleus) plus the distance from the nucleus in arcseconds \citep[as in][]{massaro11}.\\
\tablenotemark{b} Size of the aperture used for photometry.\\
\tablenotemark{c} Total counts in the photometric circle, with the background estimated from the CCD  and rescaled on the component region, in parentheses; both counts refer to the total 0.5 to 7 keV band. In particular, for the source 3CR\,124, labelled with ``*'', since there is diffuse X-ray emission surrounding the nuclear region, we estimated the local background at the same angular separation of the radio lobe/relic, within an annulus of 5$\arcsec$ inner radius and 10$\arcsec$ outer radius (see also \S~\ref{sec:sources}). \\
\tablenotemark{d} The confidence level of each detection evaluated adopting a Poisson distribution.\\
\tablenotemark{e} Fluxes extracted from the flux maps in the three energy bands (0.5-1 keV, 1-2 keV, 2-7 keV) and the total flux in the energy range 0.5-7 keV. The 1$\sigma$ uncertainties in parentheses are derived from the uncertainties on the number of counts in the source and background regions, added in quadrature.\\
\tablenotemark{f} X-ray luminosity in the range 0.5 to 7 keV, computed using the values of luminosity distance D$_L$ given in Table~\ref{tab:main}. A 1$\sigma$ uncertainty is derived from the uncertainties on the flux and it is given in parentheses.\\
\end{table*}

\subsubsection{X-ray Spectral Analysis of the stronger nuclei}
\label{sec:spectra}

We performed X-ray spectral analysis for the nuclei containing 300 or more total
counts to estimate their X-ray spectral indices,
$\alpha_X$. The 400 counts threshold adopted in previous analysis \citep[see e.g.,][]{massaro17} was lowered for this higher redshift sample.  We carried out the analysis on the level 2 event files using {\sc Xspec} version 12.9 \citep{arnaud96} and CIAO 4.9 {\sc Sherpa} version 1 \citep{freeman01} software packages, with consistent results.

We extracted the spectral data from the same 2\arcsec\ photometric aperture centered around the nuclei described above, using the
{\sc ciao} routine \texttt{specextract} while the background spectra were extracted in nearby rectangular regions, not containing obvious X-ray sources.
Then, we binned the background-subtracted spectra
to a 30 counts per bin minimum threshold to ensure the validity of $\chi^2$ statistics, and we selected the energy range 0.5-7 keV for the spectral fitting.

We fitted the photon fluxes of the four brightest sources with a simple multiplicative model, \texttt{phabs}$\times$\texttt{powerlaw} in {\sc xspec} syntax: 
a power law absorbed by the Galactic equivalent hydrogen column density, $N_{H,Gal}$ \citep[Table~\ref{tab:main},][]{kalberla05}. The normalisation parameter and the X-ray spectral
index, $\alpha_X$, were allowed to vary. Results of the fit procedure are reported in Table~\ref{tab:spec}.

We also considered the possible presence of mild pileup in \chn\ observations of the two nuclei with the highest number of counts in our sample: 3CR\,69 and 3CR\,220.2. Pileup occurs on X-ray CCDs for sources with high flux levels when two or more photons arrive within the same detector pixel within a single CCD frame integration time, and they are counted as a single photon of higher energy. 

We first produced pileup maps with the CIAO \texttt{pileup\_map} tool, which indicated a pileup fraction of $\sim$5\% and $\sim$10\% in the nuclei of 3CR\,69 and 3CR\,220.2, respectively. To constrain more precisely the pileup amount in this observations we performed a spectral fitting with {\sc Sherpa}, adding to the previous model the jdpileup model \citep{davis01}. Following the {\sc Sherpa} thread\footnote{http://cxc.harvard.edu/sherpa/ahelp/jdpileup.html} we fixed the parameters: grade zero probability, $g_0=1$, number of detection cells, $n=1$, and maximum number of photons considered for pileup in a single frame, $n_{terms}=30$. In addition, the frame time, $ftime$, and fractional exposure, $fracexp$, parameters were fixed to the values of the keywords EXPTIME and FRACEXPO of the event and ARF files. The probability of a good grade when two photons pile together, $\alpha$, and the fraction of flux falling into the pileup region, $f$, were left free to vary during the fit.

The power-law spectral index we obtained for 3CR\,69 is consistent with the 0.4 value reported in Table~\ref{tab:spec}, obtained without the pileup model, and the pileup fraction as evaluated from the \texttt{get\_pileup\_model} {\sc Sherpa} command is zero. Since the quality of the data collected for 3CR\,36 and 3CR\,119, observed to have lower flux levels, is not sufficient to allow a more detailed spectral analysis, we conclude that these sources are affected by a pileup fraction $\lesssim$5\% and that the estimate of their spectral index is not compromised.

In the case of 3CR\,220.2 the pileup fraction obtained from the spectral fitting with {\sc Sherpa} is 8.5\%, comparable with that evaluated from the pileup map and the spectral index obtained with the pileup model is reported in Table~\ref{tab:spec}. Although the two values are consistent within the 1$\sigma$ range, we note that the pileup effect tends to be more severe in this source, hardening the spectrum. The spectral fit performed with the pileup model yields estimates of the flux in the soft, medium, hard and total bands of 1.3, 1.6, 3.7, and 6.6~$\times$~10$^{-13}$~erg~cm$^{-2}$~s$^{-1}$, respectively. The total source luminosity is 5.1~$\times$~10$^{45}$~erg~s$^{-1}$.

\begin{table}
\begin{center}
\caption{Spectral Analysis of Bright Nuclei}
\label{tab:spec}
\begin{tabular}{lccccl}
\hline 
\hline    
3CR    &  $N_2$\tablenotemark{a}  &     $\alpha_X\tablenotemark{b}$      & $\chi^2$(dof)\tablenotemark{c} \\
\hline 
\noalign{\smallskip}
36     & 335(18) &  0.7(0.1)   &    5.68(7)    \\
69  & 492(22) &  0.4(0.1)    &  21.38(13)  \\                
119    & 381(20) &  0.5(0.1)   &  8.49(9)    \\           
\multirow{2}*{220.2\tablenotemark{d}}  & \multirow{2}*{683(26)\big\{} & 0.60(0.07)   &  13.7(19)  \\
                                   &                                     & $0.81^{+0.08}_{-0.25}$ & 13.2(17) \\         
\noalign{\smallskip}
\hline
\end{tabular}\\

\end{center}
\textbf{Notes.} These four nuclei have 300 or more total counts in the photometric aperture and are thus suitable for spectral analysis. \\
\tablenotemark{a} Total number of counts within a circle of radius=2\arcsec . The uncertainties given in parentheses are computed as $\sqrt{total-number-of-counts}$.\\
\tablenotemark{b} X-ray spectral index $\alpha_X$ with 1$\sigma$ uncertainties derived from the fit in parenthesis.\\
\tablenotemark{c} $\chi^2$ of the fit procedure with degrees of freedom in parentheses.\\
\tablenotemark{d} For the source 3CR\,220.2, whose observation was found to be affected by a pileup fraction of $\sim8.5\%$, we reported fit results obtained with the simple model of a power law with Galactic absorption (first row) and with the addition of the jdpileup model (second row). \\
\end{table} 

\section{Results}
\label{sec:results}

\subsection{General}
\label{sec:general}

In the current sample of 16 X-ray detected sources, three are classified as compact steep spectrum (CSS) radio sources, namely: 3CR\,119, 3CR\,173 and 3CR\,305.1 \citep{odea98}. Among the other sources, three are quasars, nine are FR\,II radio galaxies and one is an FR\,I. Radio classifications reported in Table~\ref{tab:main} are based on a literature search, with the exception of the radio galaxy 3CR\,255 for which we propose an FR\,II classification.

We detected X-ray emission for all the nuclei in the sample with a level of significance higher than 3$\sigma$. All the results of the X-ray photometry (i.e., nuclear X-ray fluxes in the three energy bands, together with their X-ray hardness ratios and luminosities) are reported in Table~\ref{tab:fluxes} (see also \S~\ref{sec:sources} for more details on each source). X-ray images are presented in Appendix~\ref{sec:appendixA} with contours  of the radio structures overlaid.

\begin{figure} 
 \subfigure
   {\includegraphics[scale=0.335, angle=0]{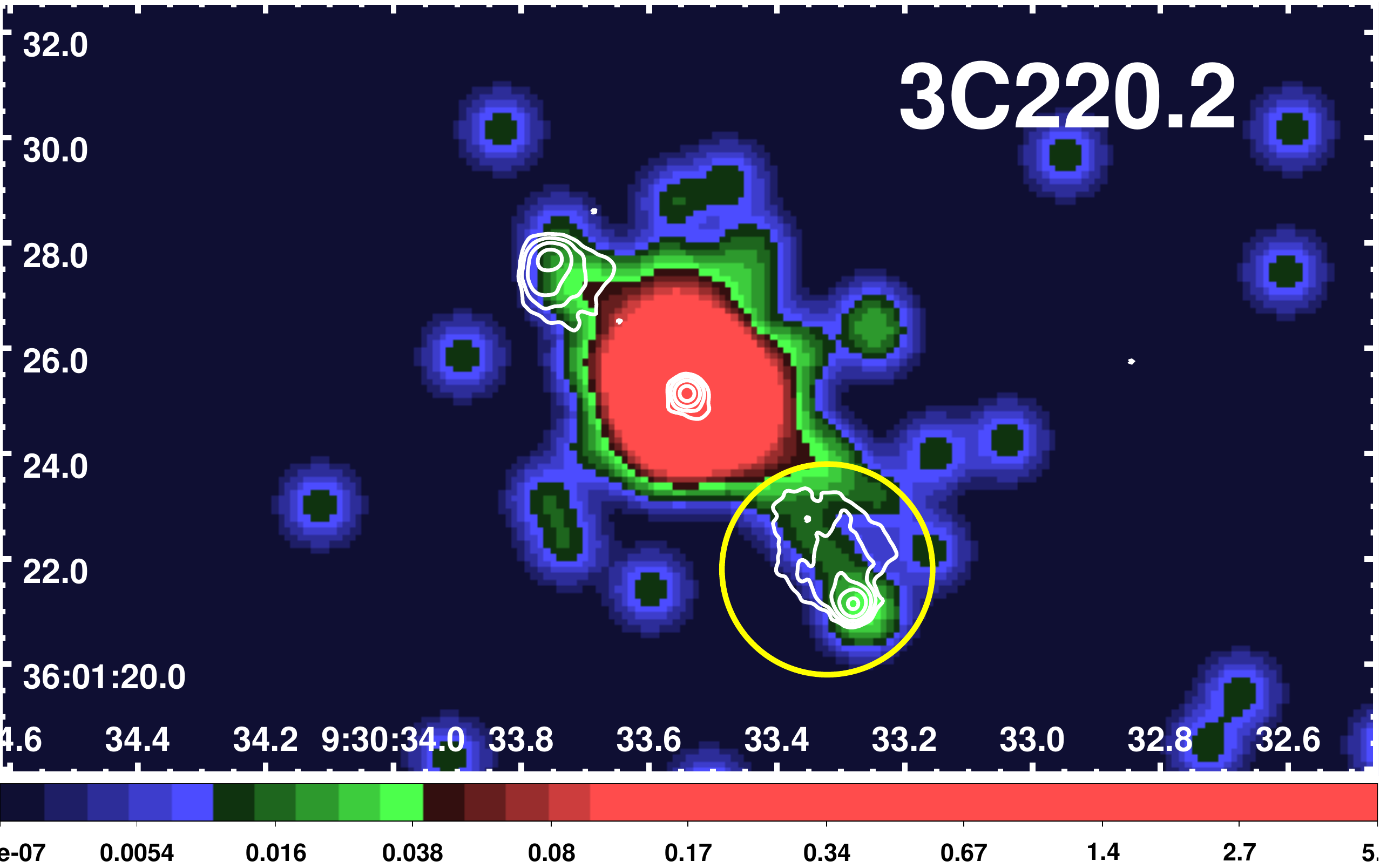}}
\hspace{5mm}
\subfigure
   {\includegraphics[scale=0.339, angle=0]{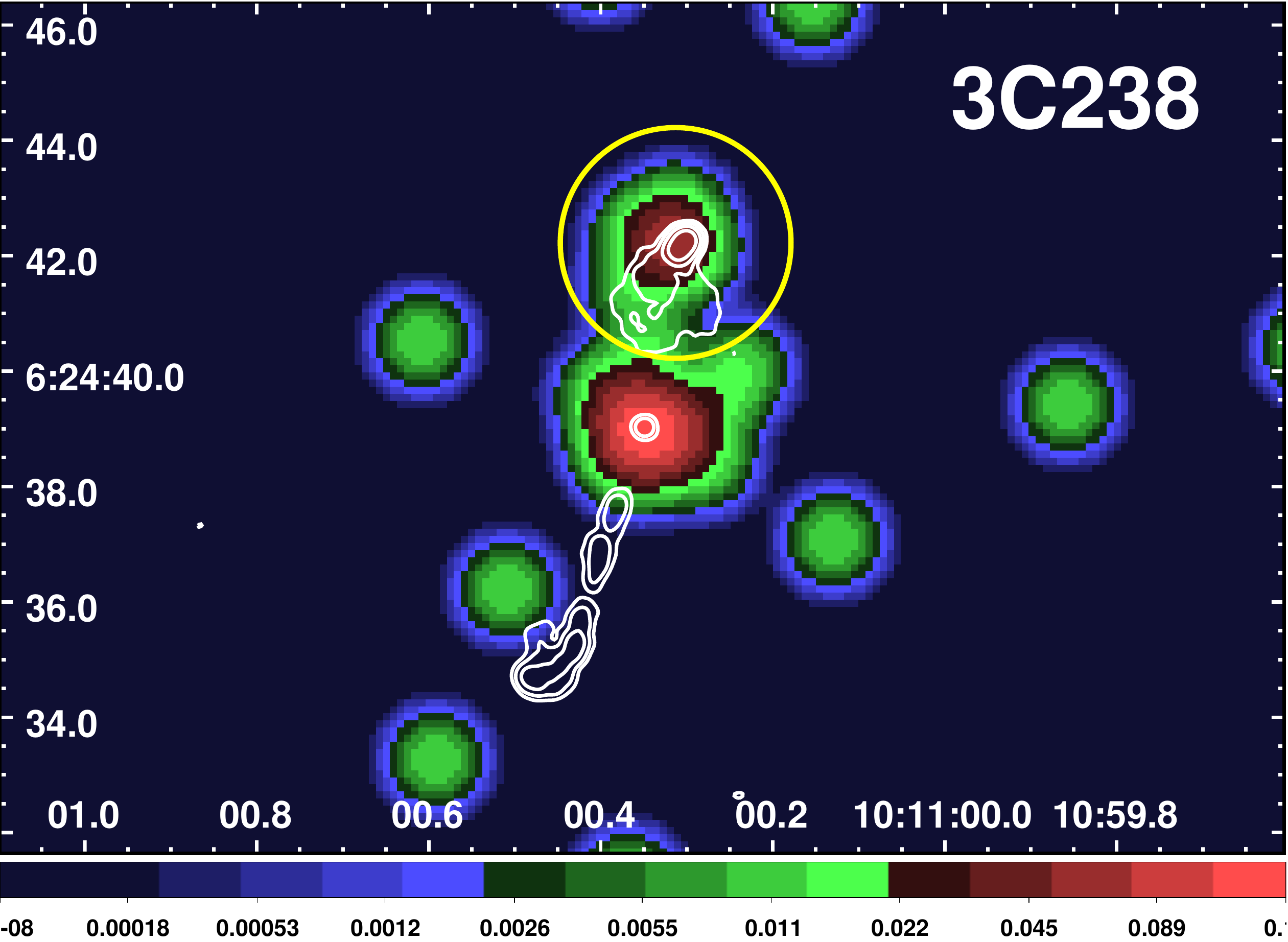}}
\caption[]{\chn\ images in units of counts/pixel (bin factor of 1/4) of 3CR\,220.2 (top) and 3CR\,238 (bottom) in the energy band 0.5-7 keV. Both images were smoothed with a Gaussian of FWHM=1\asec .0. The radio contours overlaid (white), from 8.4 GHz maps, kindly supplied by C. C. Cheung, start at 0.3 mJy/beam, increasing by factors of four.  The yellow circles are the regions used for the X-ray photometry of the detected hotspots.} 
\label{fig:hotspots}
\end{figure}

\begin{figure}
\includegraphics[scale=0.343, angle=0]{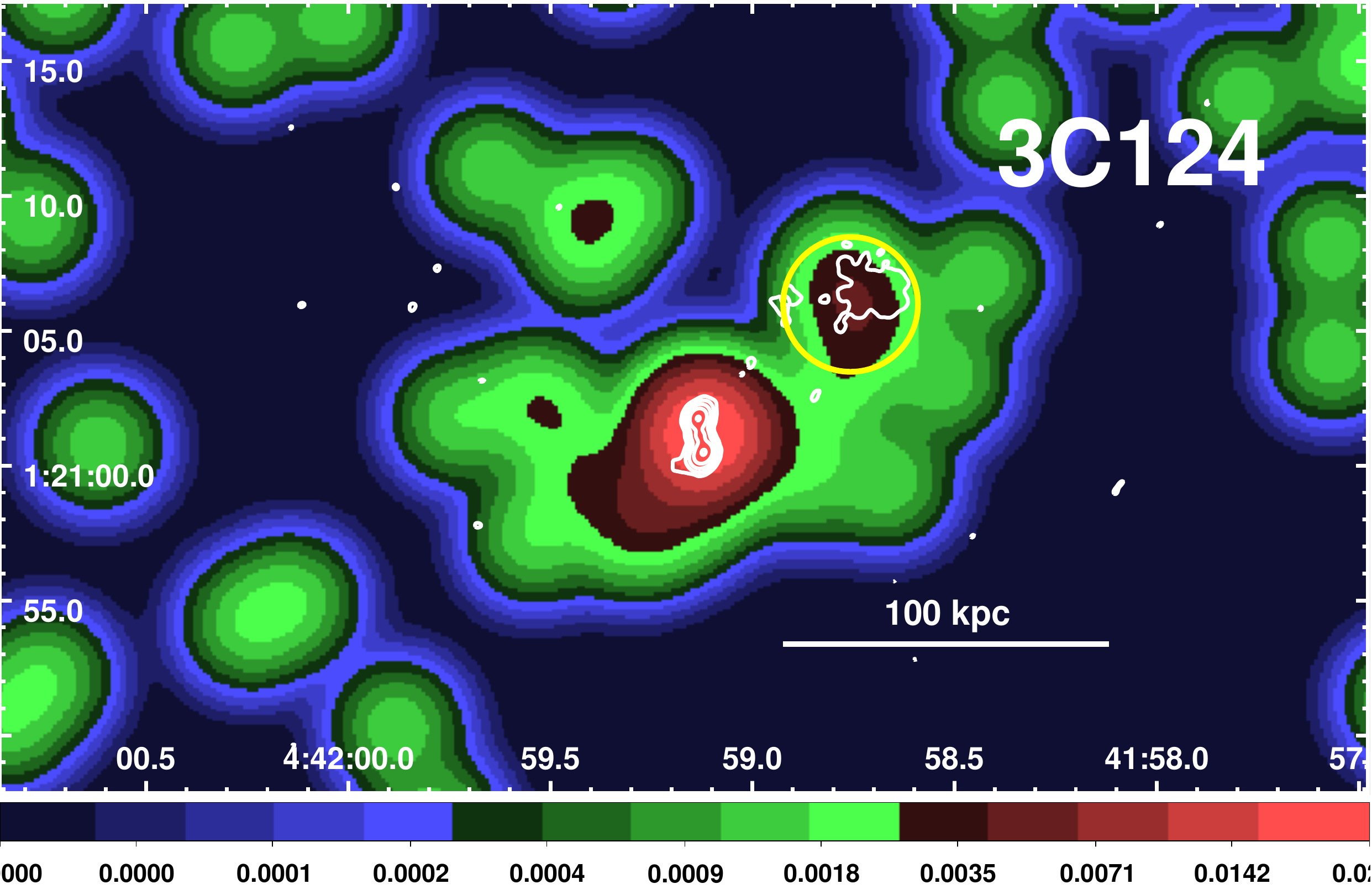}
\caption[]{The X-ray image of 3CR\,124 in units of counts/pixel for the energy range 0.5-7 keV and bin factor 1/4. The image was smoothed with a Gaussian of FWHM=3\asec .0. The radio contours (white), from an 8.4 GHz map kindly provided by C. C. Cheung, start at 0.15 mJy/beam and increase by factors of four. The NW external (relic ?) lobe is detected in the X-ray and marked with the yellow circle, and the extended emission, probably arising from the hot gas in the intergalactic medium surrounding the radio source, is clearly visible in the WNW-ESE direction.}
\label{fig:3CR124selected}
\end{figure}

\begin{figure} 
 \subfigure
   {\includegraphics[scale=0.46]{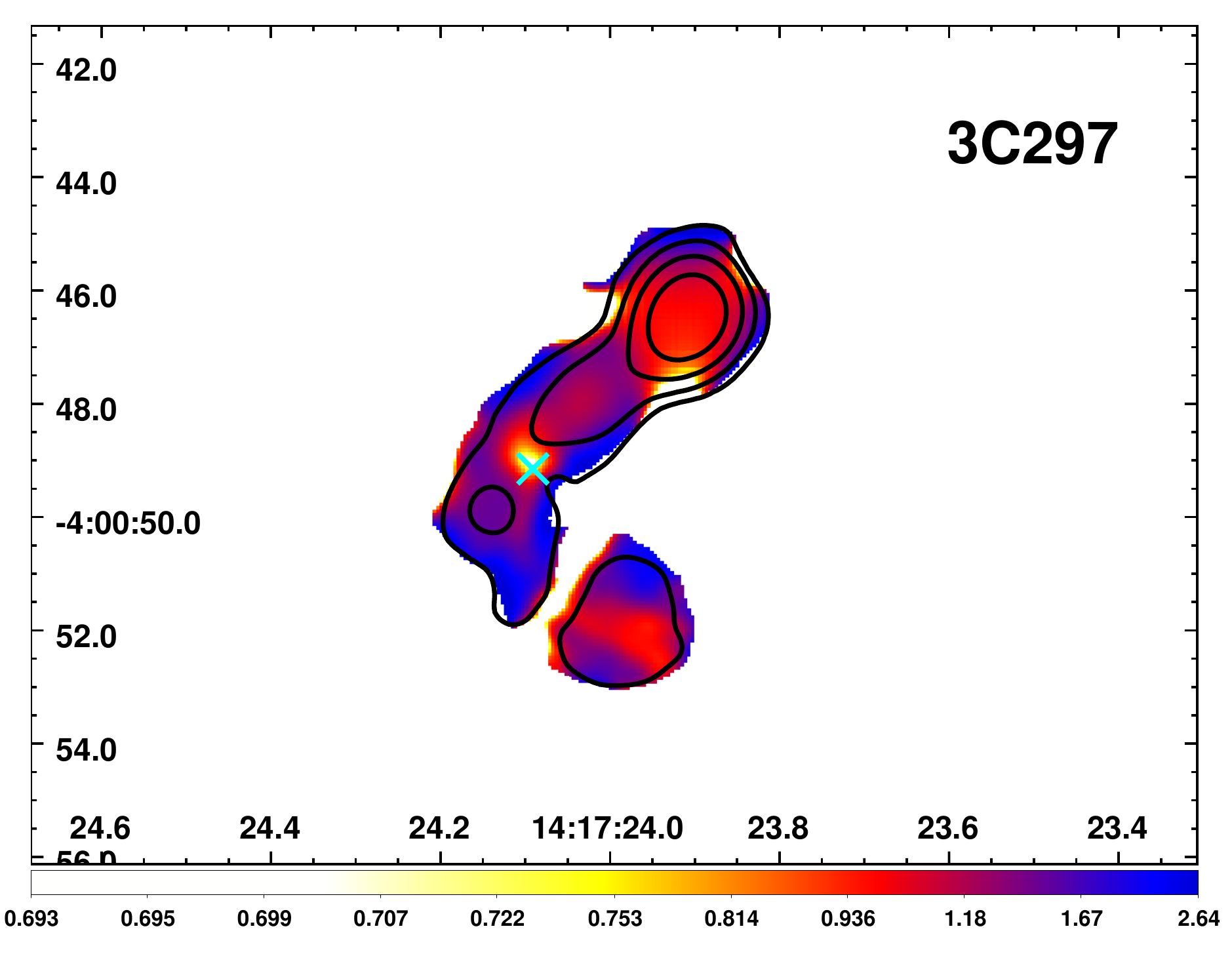}}
\hspace{5mm}
\subfigure
   {\includegraphics[scale=0.45]{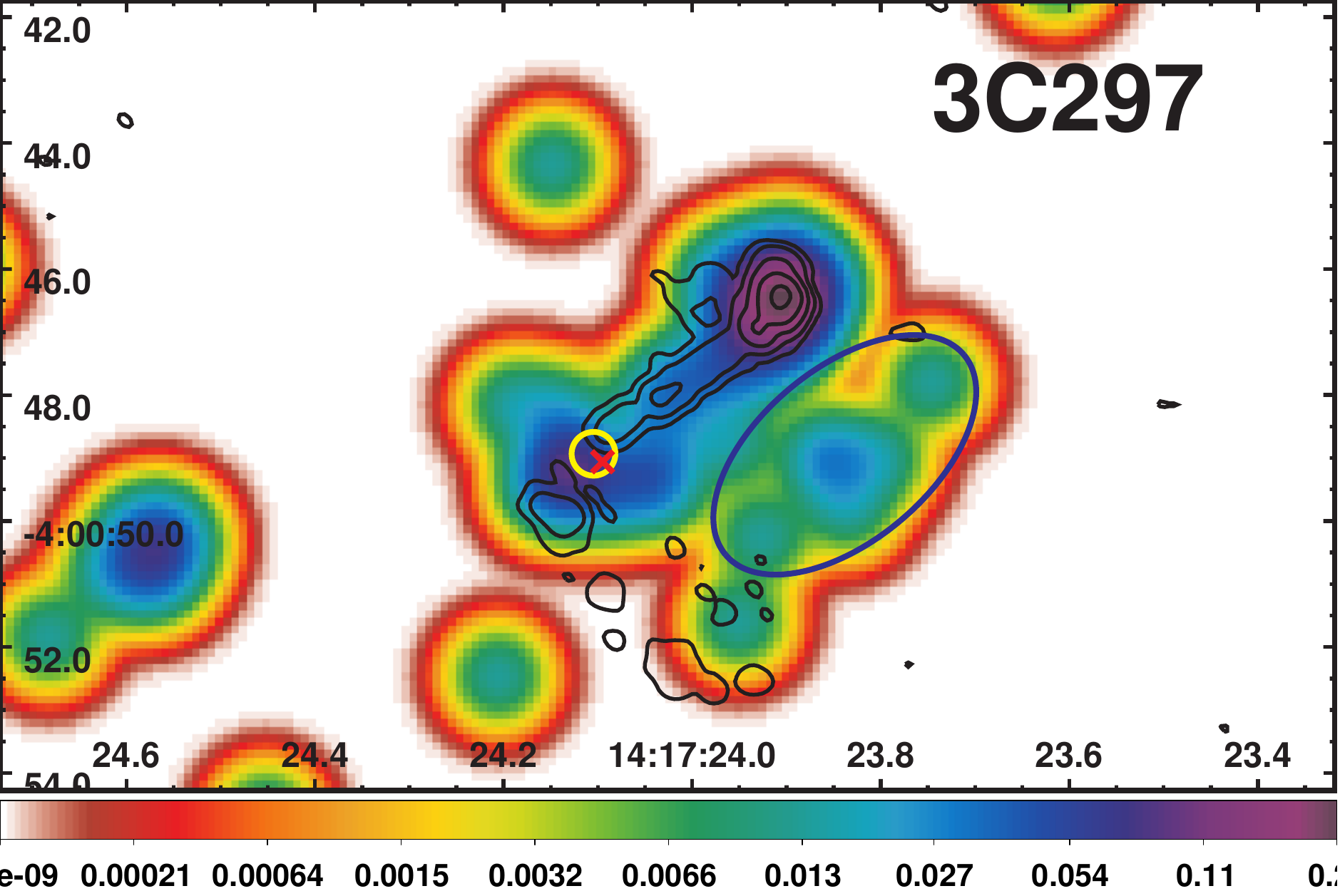}}
\caption[]{Top: spectral index map of 3CR\,297, computed in \S~\ref{sec:Radobs}, with black radio contours of the 4.9 GHz radio image superposed, starting at 2 mJy/beam and increasing by factors of four. The position of the radio core suggested from the flattening of the spectrum is marked with the cyan ``$\times$'' and shown also in the bottom image. Bottom: The X-ray image of 3CR\,297 for the energy band 0.5-7 keV and bin factor of 1/4, in units of counts/pixel. The image was smoothed with a Gaussian of FWHM=1\asec .0. The black radio contours come from an 8.4 GHz map \citep[same as in][]{hilbert16} and start at 0.08 mJy/beam, increasing by factors of six. Since the radio core position is unclear, the two images are both registered with the surface brightness peak position of the knot $k$-w\,4 (see Table~\ref{tab:jets}). The position of the host galaxy in the optical HST observation is marked with yellow circle while the position of the radio core derived from the spectral index map is marked with the red ``$\times$''. The radio enhancement visible at $\sim$1\asec\ south-east from the host galaxy is probably due to a field galaxy also observed in optical images, while the X-ray emission in the westward direction, marked with the blue ellipse, has no optical or radio counterparts and could arise from the hot gas in the intergalactic medium on one side of the source. } 
\label{fig:3CR297combo}
\end{figure}

We detected the high energy counterpart of two hotspots: one in the quasar 3CR\,220.2 and one in 3CR\,238 (see Figure~\ref{fig:hotspots}). We discovered also X-ray emission associated with a lobe in 3CR\,124 (Figure~\ref{fig:3CR124selected}), and with a jet knot in the quasar 3CR\,297 (Figure~\ref{fig:3CR297combo}). Detection significances, fluxes and luminosities of the components are reported in Table~\ref{tab:jets}.

Results of the X-ray spectral analysis performed for the brightest nuclei (i.e., 3CR\,36, 3CR\,69, 3CR\,119 and 3CR\,220.2) are reported in Table~\ref{tab:spec}: all spectra are consistent with a power-law model with Galactic absorption.

\subsection{Extended emission}
\label{sec:extended}

In Table~\ref{tab:fluxes} we also listed the total number of counts within a circular region of radius 2\asec, $N_2$, centered on the 
source radio nucleus. We used the radio position given in Table~\ref{tab:main} for those sources for which the X-ray image was not registered while for 3CR\,297 we used the radio core position suggested from the flattening of the radio spectrum (see \S~\ref{sec:sources}). As for the previous analyses of 3CR sources observed, we computed the ``Extent Ratio'' dividing $N_2$ by the number of counts in a 
region of radius 10\asec, $N_{10}$, centered on the same position, after the subtraction of background counts measured on the CCD from both values (i.e., Ext. Ratio shown in Table~\ref{tab:fluxes}). We used the ``Extent Ratio'' as a diagnostic tool
for the presence of extended X-ray emission around the source region \citep[e.g.][]{massaro10,massaro09b}, though it does not provide 
information on whether the emission is related to any extended radio structure or not.
We expect values close to unity for an unresolved (i.e., point-like) source since the
on-axis encircled energy for r=2\arcsec\ is $\approx$0.97, so there is only a small increase between 
r=2\arcsec\ and r=10\arcsec\ for an unresolved source. We found that 3 of the 16 sources in our sample show values of 
``Extent Ratio'' less than 0.5, namely: 3CR\,124, 3CR\,297, and 3CR\,300.1; indicating the possible presence of diffuse X-ray emission. For 3CR\,300.1, the indication of the presence 
of extended X-ray emission, given by the low value of ``Extent Ratio'', is not reliable because its value 
is affected by the presence of an X-ray point source at $\sim$7\asec distance from its nucleus.

All sources showing extended X-ray emission are at $z>$1.0, thus the combination of their radio projected size, close 
to the limit of \chn\ resolution, and the short exposures of snapshot observations do not always 
allow us to distinguish between X-ray extended emission arising from the hot intergalactic medium (IGM) in their large-scale environment and the X-ray radiation related to their large-scale radio structure (i.e., as for example IC/CMB in their radio lobes). 
We interpret the diffuse X-ray emission as due to the high energy 
counterpart of the extended radio structure when the peaks of X-ray surface brightness appear coincident
with those of the radio intensity. Otherwise we conclude that it could be due to the presence of thermal emitting gas in the 
IGM on a scale of several tens to hundreds kpc. Radio images in the MHz band, together with 
X-ray spectroscopic observations, are necessary to clearly disentangle the two above scenarios \citep{croston05}. An example is the radio galaxy 3CR\,124 (see Fig. \ref{fig:3CR124selected}) with an ``Extent ratio'' of 0.49. In this source we detected both extended X-ray 
emission surrounding the radio source, on tens kpc scale and X-ray diffuse emission coincident with the external lobe, 
asymmetrically located with respect to the main radio axis (detection significance and fluxes 
of the lobe are all reported in Table~\ref{tab:jets}). Both emission regions contribute to the low value of ``Extent Ratio''.

We detected X-ray emission with a brightness peak to the west side of the quasar 3CR\,297 
(see Figure~\ref{fig:3CR297combo}). Since the radio core position of this source is unclear, radio and X-ray images are registered with the surface brightness peak position of the knot $k$-w\,4 (see Table~\ref{tab:jets}). The presence of diffuse X-ray emission in the westward direction is not related to any radio structure 
and we did not find an obvious optical counterpart in HST images, thus we interpreted it as X-ray radiation 
arising from the IGM. A detailed description of this peculiar source is given in \S~\ref{sec:sources}.

None of the 16 sources in our sample has been reported in the literature as a member of a galaxy cluster 
or group, although we found that, based on the literature search reported in \S~\ref{sec:sources}, 3CR\,220.2, 3CR\,255, and 3CR\,300.1 could lie in large-scale (i.e., 100 kpc) galaxy-rich environments. For these sources we searched for possible X-ray emission due to the hot gas of the intracluster medium around 
their radio structures \citep[see also][]{massaro17}. 
We measured the total number of photons in a circular 
region containing the entire extent of the radio source and we subtracted the X-ray counts within circular 
regions of 2\asec, corresponding to the radio nucleus, as well as that of point-like sources (e.g. hotspots and/or background/foreground objects) lying within the same region. Assuming a Poisson distribution for the 
background events, we computed the probability of obtaining the measured value. We did not find any X-ray excess above 3$\sigma$ significance on a few hundreds kpc scale environment. 

\subsection{Source details}
\label{sec:sources}

\underline{{ 3CR\,27}} is a classical FR\,II radio galaxy at $z$=0.184. From an optical perspective it is an elliptical High Excitation Radio Galaxy (HERG), as confirmed by the spectrum used to update its optical identification after the last revision of the 3CR catalog \citep{hiltner91}. It is located in a region of high optical obscuration at low Galactic latitude (i.e., $b$=5$^\circ$.5). With the current radio map at 1.4 GHz, it was not possible to locate the radio core. However, its X-ray nucleus was probably detected with a total of 63 counts in a region of radius $\sim$2\asec\ centered on the celestial coordinates reported in Table~\ref{tab:main} \citep{hiltner91} and located between the two lobes of the radio galaxy (see Figure~\ref{fig:3CR27}). The SW hotspot has a flux density $\sim$4 times greater than the northern one at 1.4 GHz. Salter \& Haslam (1980) reported the detection of a possible over density of sources in the radio observations at 408 MHz and at 2.7 GHz 
over an area of 3.1$^{\circ}\times$1.3$^{\circ}$ surrounding 3CR\,27, without any information on its redshift. However given the $z$ estimate available in the literature this corresponds to a linear size $\sim$35 Mpc, making the possible detection of a galaxy rich environment around 3CR\,27 unlikely.
In addition extended X-ray emission arising
from the intergalactic medium on a hundreds kpc
scale was not detected in our \chn\ snapshot observation.\\

\underline{{ 3CR\,36}} is classified as an FR\,II HERG at $z$=1.301 \citep{hewitt91,jackson97}. In the 8.4 GHz map, its angular extension is $\sim9\asec$, corresponding to a projected size of less than 80 kpc, with a dominant knotty jet in the north. In the HST optical image, 3CR\,36 appears as a compact galaxy, and only a central nuclear component is detected \citep{mccarthy97}. We detected the nucleus also in our \chn\ snapshot survey (see Figure~\ref{fig:3CR36}) and the total 335 X-ray photons observed in the nuclear region allowed us to perform a spectral analysis, which yielded a spectral index $\alpha_X$=0.7 (see Table~\ref{tab:spec}).\\

\underline{{ 3CR\,69}} is an FR\,II radio galaxy, optically classified as HERG, at $z$=0.458 \citep{hiltner91}. Since the source lies in a heavily obscured region at Galactic latitude $b$=-0$^\circ$.9, its optical identification was made only after the last revision of the 3CR catalog. The radio image at 8.4 GHz shows two extended lobes and a fainter nucleus that was the only X-ray detection (see Figure~\ref{fig:3CR69}). The spectral analysis of the nucleus yielded a spectral index $\alpha_X$=0.4, as reported in Table~\ref{tab:spec}, and the pileup fraction estimate is $\sim$5\%. \\

\underline{{ 3CR\,119}} is a CSS radio source \citep{odea98} and its updated redshift is $z$=1.023 \citep{eracleous94}. The source, which in our VLA 8.4 GHz map is a compact 0\asec.5 structure ($\sim$5 kpc at the source redshift),  shows a very complex radio morphology on the 1-kpc scale observed with the VLBI at 5 GHz: the nucleus was identified with a weak compact ($<$ 0\asec.003) component with flat spectrum, while the extended structure has a spiral-like form with two knots along the jet \citep{fanti86}. This distorted morphology was interpreted as the result of the source interaction with dense clouds in its environment and this idea was supported by the presence of a large gradient of the rotation measure (i.e., 2300 rad/m$^2$/mas in the 8.4-GHz band) and strong depolarisation between 8.4 and 5 GHz \citep{nan99,mantovani10}. The X-ray snapshot observation revealed point-like emission on a scale of $\sim2$\asec, corresponding to $\sim$15 kpc at the source redshift, shown in Figure~\ref{fig:3CR119}. The spectral analysis of its nucleus yielded a value of $\alpha_X$=0.5 (see also Table~\ref{tab:spec}).\\

\underline{{ 3CR\,124}} is an FR\,II radio galaxy at $z$=1.083 \citep{hewitt91}. Optically, it is classified as a HERG \citep{jackson97}. The 8.4 GHz image shows a small scale compact double-lobe structure, with a fainter central core and angular extension of 2\asec.5 , corresponding to about 20 kpc. However, 3CR\,124 also shows additional, more extended, asymmetric radio emission misaligned by $\sim45^\circ$ measured in the west direction from the line connecting the other two, separated by $\sim$6\asec\ (see Figure~\ref{fig:3CR124}), suggestive of a relic lobe \citep{mccarthy89,dunlop93}. The optical galaxy observed with HST is more extended than the radio source, it has a highly curved morphology and shows an emission-line region which is very closely aligned with the radio one \citep{mccarthy97,privon08}. The two radio components aligned in the NS direction can be interpreted as the two sides of an outflow, because a velocity gradient aligned with the radio axis was measured in the form of extended  [O III] $\lambda$5007 emission regions \citep{shih13}. 

The value of ``Extent Ratio'' computed for this source, 0.49$\pm$0.09 (see Table~\ref{tab:fluxes}), suggests the presence of diffuse emission around the nuclear region of the source. Examining the image more closely, we detect 22 photons in the 2\asec-10\asec\ radial region, where we expect 5 background photons giving a detection of extended emission of more than 6$\sigma$. In this region the total luminosity is 1.1$\pm$0.2~$\times$~10$^{44}$~erg~s$^{-1}$. 

We found an excess of X-ray photons with respect to the local background with more than 3$\sigma$ level of significance in a region of radius 2.5\asec (see Table~\ref{tab:jets}). This is spatially coincident with the lobe/relic radio emission appearing in the NE direction (see Figure~\ref{fig:3CR124selected}). The local background was estimated at the same angular separation of the radio lobe/relic, thus using an annulus of 5$\arcsec$ inner radius and 10$\arcsec$ outer radius. However, the low number of X-ray photons does not allow us to distinguish if such emission is due to a fluctuation of the diffuse X-ray radiation that appears surrounding the nucleus also in the SE direction or due to emission arising from particle accelerated in the NE radio structure. 

We could favour the latter scenario because (i) the excess of X-ray photons, spatially coincident with the NE radio structure, is $\sim$2 times larger than the number of counts in the SE direction, at the same angular separation and measured over the same area, and since (ii) it lies at larger distance in the NE direction than the rest of the diffuse X-ray emission in the southern side.

Nuclear X-ray emission was also detected by \chn\ (see Figure~\ref{fig:3CR124}).\\

\underline{{ 3CR\,173}} is a CSS at $z$=1.035, optically classified as a HERG \citep{jackson97}. A double structure is visible in the radio image at 4.9 GHz, with the two peaks of radio surface brightness separated by only $\sim$2\asec, correspondent to $\sim$15 kpc at the source redshift (see Figure~\ref{fig:3CR173}). The core is the dominant south-west component, characterised by a flattening of the integrated radio spectrum near 38 MHz \citep{tyul00} and it is the only X-ray detection in this source.\\

\underline{{ 3CR\,194}}, at $z$=1.184 \citep{strom90}, was optically classified as a HERG \citep{jackson97} and it is an FR\,II radio galaxy with a double hotspot in the NW lobe (see Figure~\ref{fig:3CR194}). A high-sensitivity VLA image at 8.4 GHz showed that the two radio lobes are asymmetrically located with respect to the radio core and the north-western lobe does not extend along the radio source axis, but is almost perpendicular to it \citep{fernini14}. Weak radio emission in the EW direction was also observed at 5 GHz, giving an irregular quadrupolar shape and suggesting a movement of the source in the ambient medium in a westward direction \citep{strom90}. The optical galaxy in HST images is elongated North-South and composed of two components separated by 2\asec \citep{djorgovski88,mccarthy97}. 3CR\,194 displays rotation measures (in high resolution VLA observations at 1.4, 4.9 and 8.4 GHz) in excess of 1000 rad m$^{- 2}$  with an enhancement coincident with the NW hotspot \citep{taylor92}: the excess was interpreted as due to the compression of external gas and magnetic field by the terminal shock that created the hotspot. This suggests the presence of high density hot gas surrounding the galaxy, however lacking an X-ray detection. With 124 counts, the X-ray emission of the nuclear region of this galaxy was clearly observed by \chn\.\\

\underline{{ 3CR\,208.1}} is a quasar (QSO). The optical counterpart was first identified with a N-type galaxy at $z$=1.02 \citep{spinrad85}. Gravitational amplification due to a foreground source $\sim3\asec$ SE at $z$= 0.159 was found to brighten the QSO by several tenths of a magnitude and to increase its radio flux by a factor of 1.5 \citep{lefevre90}. The optical image of 3CR\,208.1 has a compact nucleus and a secondary component 0.6\asec\ NE \citep{mccarthy97} while the radio image at 14.9 GHz shows a lobe 5\asec\ NW with a curved jet. The only feature detected by \chn\ is the radio core (see Figure~\ref{fig:3CR208.1}).\\

\underline{{ 3CR\,220.2}} is a QSO at $z$=1.157 \citep{spinrad85}. In the radio image at 4.9 GHz, both lobes, separated by $\sim$60 kpc, show a hotspot in the outer region. The peak of the radio surface brightness in the southern hotspot is coincident with a feature visible in the HST optical image, but also with the peak of the X-ray emission detected and designated as $h$-s 5 in Table~\ref{tab:jets}. Two X-ray photons are detected in the radio NE lobe/hotspot region. However, given the presence of X-ray photons at similar angular separation from the radio core with no radio counterparts, we cannot exclude that those in the northern lobe/hotspot are fluctuations of the local background. The X-ray nuclear emission is consistent with an unresolved point source, despite the presence of an elongation of the X-ray emission along the radio axis visible in Figure~\ref{fig:hotspots} and Figure~\ref{fig:3CR220.2}. 

In IR and optical images the source is aligned with a group of large spiral galaxies \citep{hilbert16}. Based on the photometric redshift estimations of the Sloan Digital Sky Survey \citep{sdss13}, these objects are probably foreground galaxies. However, we searched for extended X-ray emission on a scale of hundreds kpc in the field of 3CR\,220.2 finding no neat detection.

The nucleus of 3CR\,220.2 is the brightest one within the sample (i.e., $N_2$=683) and it has a spectral index $\alpha_X$=0.6 (see Table~\ref{tab:spec}). The pileup fraction obtained for this source is $\sim$8.5\% but the spectral index obtained using a pileup model (i.e., $\alpha_X=0.81^{+0.08}_{-0.25}$) is consistent with 0.6. \\

\underline{{ 3CR\,222}} is known to be the most distant FR\,I radio galaxy of the 3CR catalog to date, having $z$=1.339 \citep{heckman94}, although there is no published spectrum in the literature. It was classified as an ultra steep spectrum radio source, since it has a radio spectral index $\alpha_R$=1.52 in the frequency range from 10 MHz to 1 GHz \citep{roland82}. 3CR\,222 was described as a compact double radio source, dominated by the nuclear emission at 1.4 GHz and with steep-spectrum emission coming from a region separated by 1\asec .2 to the south \citep{law-green95, strom90}. The optical identification is the W member of an apparent faint quartet \citep{djorgovski88}. The X-ray nucleus is detected at 3$\sigma$ level of significance with only three photons observed by \chn\ in a circular region of radius 2\asec\ (see Figure~\ref{fig:3CR222}).\\

\underline{{ 3CR\,230}} is an FR\,II radio galaxy at $z$=1.487 with a double hotspot at the end of the northern lobe \citep{hewitt91}. It is optically classified as a HERG \citep{jackson97}. The radio image at 8.4 GHz shows two knots along the southern jet: it was claimed that the radio core position is coincident with the northern one, at R.A. = $9^{h}51^{m}58^{s}.90$ and $\delta$ = $-$ $00^{\circ}01'27''.87$ (J2000.0) \citep{steinbring11}. However, the comparison with the HST \citep{hilbert16} and WISE \citep{wright10} images of the sources shows that the optical nucleus is 0\asec .5 North of this position. Moreover, since the shift that should be applied to register the X-ray image with the radio core identified by Steinbring is $\sim$1\asec .5, more than twice the average shift imposed to the other sources, we used the unregistered image (see Figure~\ref{fig:3CR230}).

By studying the relationship between source outflow and star formation in this galaxy, Steinbring noticed that the asymmetry of the 3CR\,230 radio structure is mirrored in the optical surface brightness and in the strong turbulence in the direction of the shorter radio arm to the south-west. The optical galaxy runs along a central spine SE to NW aligned with the radio lobes and ends in broader tails, while the observed linear structures due to star formation are possibly induced by the jet \citep{hilbert16}. 3CR\,230 is identified as a possible cluster candidate according to the red sequence method \citep{kotyla16}. The nucleus is the only X-ray detection although its position is uncertain.\\

\underline{{ 3CR\,238}} is a HERG with a classical FR\,II morphology at $z$=1.405 \citep{jackson97}. It could be affected by gravitational lensing by the Abell cluster A949 with an estimated redshift $z$=0.142 along its line of sight \citep{lefevre88bis}. However, the core of this galaxy cluster is outside the field of view of the \chn\ observation. The optical observation made with the HST shows a compact nucleus and faint extensions in the NS direction \citep{mccarthy97}. In the 8.4 GHz image, the southern jet shows two elongated knots separated by 1\asec\ and 2\asec\ from the nucleus while the northern knot has a wider structure that is also detected in the X-ray (see Table~\ref{tab:jets} and Figure~\ref{fig:hotspots}). The offset of $\sim0\asec.2$ between the peak of X-ray and radio emission is within the \chn\ position error \citep[see e.g.,][]{massaro11}. We also checked the HST field and we did not find any optical source that could be associated with this X-ray emission. The ``Extent Ratio'' value computed for this source, as reported in Table~\ref{tab:fluxes}, is $<$0.9 and it indicates the presence of extended emission ascribed to the X-ray radiation associated with the hotspot (see Figure~\ref{fig:3CR238}).\\

   \underline{{ 3CR\,255}} is a $z$=1.355 radio galaxy with two radio components separated by 10 kpc \citep{giraud90}. It is optically classified as a HERG \citep{jackson97}. The 8.4 GHz map indicates a weak signal at the centre of the host galaxy, as well as a brighter, compact radio source to the southeast (see Figure~\ref{fig:3CR255}). The northern radio component is located at the position of the optical galaxy, however more data are necessary to verify if it is compact and if it has a flat radio spectrum, as expected for the core of radio galaxies. In the optical band, the source appears elongated on 1\asec\ scale and an over density of objects with optical magnitude in the R band between 22.5 and 24 was found in a 62\asec$\times$62\asec\ area around the target \citep{giraud90}. This suggests the presence of a group or cluster of galaxies on the scale of $\sim$500 kpc. An IR HST image with $\sim$ 0\asec .1 resolution shows several extended sources in the vicinity of 3CR\,255 \citep{hilbert16}. In the NASA/IPAC Extragalactic Database these objects are all catalogued as infrared sources with no redshift estimation. All of them have a clear optical counterpart, suggesting that active star formation is ongoing \citep{hilbert16}. The possible suggestion of the presence of a small group of galaxies in the field of 3CR\,255 comes also from the red sequence method \citep{kotyla16}. However we did not detect extended X-ray emission in the hundreds kpc environment of this source. 3CR\,255 lies entirely in a region of 2\asec\ radius in which \chn\ observed only five X-ray photons. \\

\underline{{ 3CR\,297}} is a quasar with a peculiar radio morphology, lying at $z$=1.406 \citep{spinrad85}. The emission observed in the 8.4 GHz and 4.85 GHz maps, computed in \S~\ref{sec:Radobs}, is elongated and curved with a knotty enhancement at the end of the northern jet (see Figure~\ref{fig:3CR297combo}). In both radio images the emission to the south spreads over a much wider area, with a stronger concentration of signal centered on a double elongated source approximately 8 kpc (i.e., $\sim$1\asec) in projected distance from the host galaxy. The optical image shows two components separated by 0\asec .3 \citep{mccarthy97}. Recent HST IR and optical images suggest the presence of ongoing merger activity in the area \citep{hilbert16}. 

X-ray emission surrounding the radio jet is extended and there are 33 X-ray photons coincident with the position of the peak of the radio intensity of the jet knot in the NW (see Figure~\ref{fig:3CR297}). Its detection, X-ray fluxes and luminosity are reported in Table~\ref{tab:jets}. The precise location of the 3CR\,297 nucleus in the radio maps available to us was uncertain, thus we decided to perform the astrometric registration using the radio position of the jet knot (i.e., R.A. = $14^{h}17^{m}23^{s}.905$, $\delta$ = $-$ $04^{d}00^{m}46^{s}.42$, J2000.0), assuming that the peak of X-ray surface brightness coincides with the radio peak. We then estimated the radio core position, indicated by the flattening of the radio power-law spectrum, using the spectral index map computed in \S~\ref{sec:Radobs}. As shown in Figure~\ref{fig:3CR297combo}, the minimum value of spectral index, $\alpha_R$=0.68, is reached at: R.A. = $14^{h}17^{m}24^{s}.095$, $\delta$ = $-$ $04^{d}00^{m}49^{s}.06$ (J2000.0), almost coincident with the optical host galaxy. We used also the HST observation to compare the morphology of the source in the radio, X-ray and optical bands.

X-ray diffuse emission was observed in a 2\asec$\times$4\asec\ region west of the source with six X-ray counts and a total luminosity of 6$\pm$2~$\times$~10$^{43}$~erg~s$^{-1}$: since no optical counterpart was found in the HST image, this emission could be due to hot gas in the large scale environment. \\

\underline{{ 3CR\,300.1}} is a classical FR\,II HERG at $z$=1.159 \citep{djorgovski88}, with two lobes separated by $\sim$10\asec\ (i.e., $\sim$80 kpc at the source redshift) and a central core fainter than the lobes in the 14.9 GHz map \citep{jackson97}. The radio core is not detected in the 8.4 GHz radio image. In HST IR images, the galaxy appears as an extended irregular source with diffuse emission that suggests the presence of active star formation regions \citep{hilbert16}. 

Kotyla et al. (2016) reported a possible presence of a  source over density surrounding 3CR\,300.1 and a possible cluster/group detection using ``red sequence'' method that does not appear clearly evident to us, however we checked for the presence of diffuse X-ray emission on tens of kpc scale in our \chn\ snapshot observation. At a projected distance of $\sim$7\asec\ in the SW direction, we detected an X-ray point source which has an IR counterpart in HST image (see Figure~\ref{fig:3CR300.1}). The value of  ``Extent Ratio'' computed in Table~\ref{tab:fluxes} indicates the presence of extended emission around its nucleus but we did not find a relevant detection in the annular region surrounding the radio source, if we exclude the background/foreground source. 

The two radio lobes are detected with more than 2$\sigma$ level of significance using a local background. This was estimated in an annulus of 3\asec\ inner radius and 8\asec\ outer radius. \\

\underline{{ 3CR\,305.1}} at $z$=1.132 \citep{spinrad85}, is catalogued as a CSS \citep{odea98} and optically it is the only Low Excitation Radio Galaxy (LERG) in our sample. The optical galaxy has a complex morphology suggestive of on-going merger activity \citep{hilbert16}: the optical emission can be seen over more than 2\asec\ with a curved S-shaped structure in the inner 1\asec\ radius and a close agreement between optical and radio orientation \citep{mccarthy97}. Radio images at 8.4 GHz show the northern jet with more elongated emission and a southern circular lobe. In the 15 GHz radio map the northern jet shows three knots: the southernmost has a radio spectral index of 0.6, and it is highly polarised (60\%), therefore unlikely to be the core \citep{vanbreugel92}. No nuclear emission is detectable in the radio maps available to us, ranging from 1.5 GHz to 22.5 GHz. Thus the registration was not performed. This source is detected with 16 X-ray photons in our \chn\ image (see Figure~\ref{fig:3CR305.1}). 

\section{Summary and conclusions}
\label{sec:summary}

We presented new \chn\ X-ray observations of 16 radio sources listed in the Third Cambridge Revised (3CR) catalog, observed during \chn\ Cycle 17. The current sample lists three compact steep spectrum (CSS) sources, three quasars, nine FR\,II radio galaxies, and the most distant FR\,I radio galaxy within the 3CR extragalactic catalog. Fourteen targets lie at $z$ in the range 1.0-1.5, plus 3CR\,27 and 3CR\,69, which lie at $z$=0.184 and $z$=0.458, respectively. 

Thanks to the \chn\ snapshot survey, all 3CR extragalactic sources with $z<$1.5 have at least an exploratory X-ray observation available to the astronomical community, enabling us to: {\it i)} search for X-ray emission from jet knots, hotspots and lobes in radio sources, and {\it ii)} investigate the high energy emission arising from the intergalactic medium in the environments of quasars and radio galaxies, at different scales. 

Here, the basic source parameters for
the newly acquired data are presented. We built flux maps for all the X-ray observations in three energy ranges (0.5--1 keV, 1--2 keV and 2--7 keV) and
we provide basic parameters including: net counts, fluxes and luminosity, for the nuclei and other radio structures detected (i.e., jet knots, hotspots, lobes). In addition, we performed X-ray spectral analysis of the four brightest nuclei, finding consistency with a simple power-law spectrum with Galactic absorption.

The three CSS radio sources in our sample appear as point-like X-ray sources \citep[see][for a recent discussion of X-ray observations of CSS radio sources]{odea17}.

We discovered the X-ray counterparts of two radio hotspots one in 3CR\,220.2 and the other in 3CR\,238) and of a lobe/relic in 3CR\,124. Moreover, we discovered extended X-ray emission around the nuclear regions of 3CR\,124 and 3CR\,297 on the scale of $\sim$10\asec and $\sim$5\asec, corresponding to 85 and 45 kpc, respectively. In the quasar 3CR\,297 we discovered diffuse X-ray emission, of still unknown nature, along and parallel to the main radio axis. 

Finally, a table summarising the state-of-the-art of the X-ray (i.e., \chn\ and {\it XMM-Newton}) observations of 3CR extragalactic sources carried out to date is also reported at the end of the present manuscript (see Appendix~\ref{sec:appendixB}). \chn\ detections are based on both our previous and current analysis and represent an update with respect to previous works, while those regarding {\it XMM-Newton} are only based on literature search carried out to date.

\acknowledgments

We thank the anonymous referee for useful comments that led to improvements in the paper.
We are grateful to C. C. Cheung for providing several radio images of the 3CR sources\footnote{http://www.slac.stanford.edu/$\sim$teddy/vla3cr/vla3cr.html} while the remaining ones were downloaded from the NVAS\footnote{http://archive.nrao.edu/nvas/} (NRAO VLA Archive Survey), NED\footnote{http://ned.ipac.caltech.edu/} (NASA/IPAC Extragalactic Database) and from the DRAGN webpage\footnote{http://www.jb.man.ac.uk/atlas/}.
This investigation is supported by the National Aeronautics and Space Administration (NASA) grants GO4-15097X, GO6-17081X and AR6-17012X.  G.R.T. acknowledges support from the NASA through the Einstein Postdoctoral Fellowship Award Number PF-150128, issued by the Chandra X-ray Center, which is operated by the Smithsonian Astrophysical Observatory (SAO) for and on behalf of NASA under contract NAS8-03060. 
The National Radio Astronomy Observatory is operated by Associated Universities, Inc., under contract with the National Science Foundation.
This research has made use of data obtained from the High-Energy Astrophysics Science Archive Research Center (HEASARC) provided by NASA's Goddard Space Flight Center; 
the SIMBAD database operated at CDS, Strasbourg, France; the NED operated by the Jet Propulsion Laboratory, California Institute of Technology, under contract with the NASA.
TOPCAT\footnote{\underline{http://www.star.bris.ac.uk/$\sim$mbt/topcat/}} 
\citep{taylor05}, and SAOImage DS9, developed by the SAO, were extensively used in this work for the preparation and manipulation of the tabular data and images.\\

{Facilities:} \facility{VLA}, \facility{CXO (ACIS)}


\appendix
\section{A. Images of the sources}
\label{sec:appendixA}
Although for many radio sources in our sample the X-ray data are comprised of
rather few counts, the radio morphologies are shown here via contour
diagrams which are superposed on X-ray images. The \chn\ images shown are in units of counts/pixel 
in the 0.5-7 keV energy range. All the X-ray images were re-gridded with a bin factor 1/4 to obtain a common pixel size of 0.123\asec, and
 then smoothed with a Gaussian function. The full width half maximum (FWHM) of the
Gaussian smoothing function is given in the figure captions, 
as well as the increasing factors of the radio contours overlaid.
Figures appear different from each other because of the
wide range in angular size of the radio sources and it is worth noting that 3CR\,27 and 3CR\,69
are in a different redshift range than the other sources.

\begin{figure}
\centering
\includegraphics[scale=0.4, angle=-90]{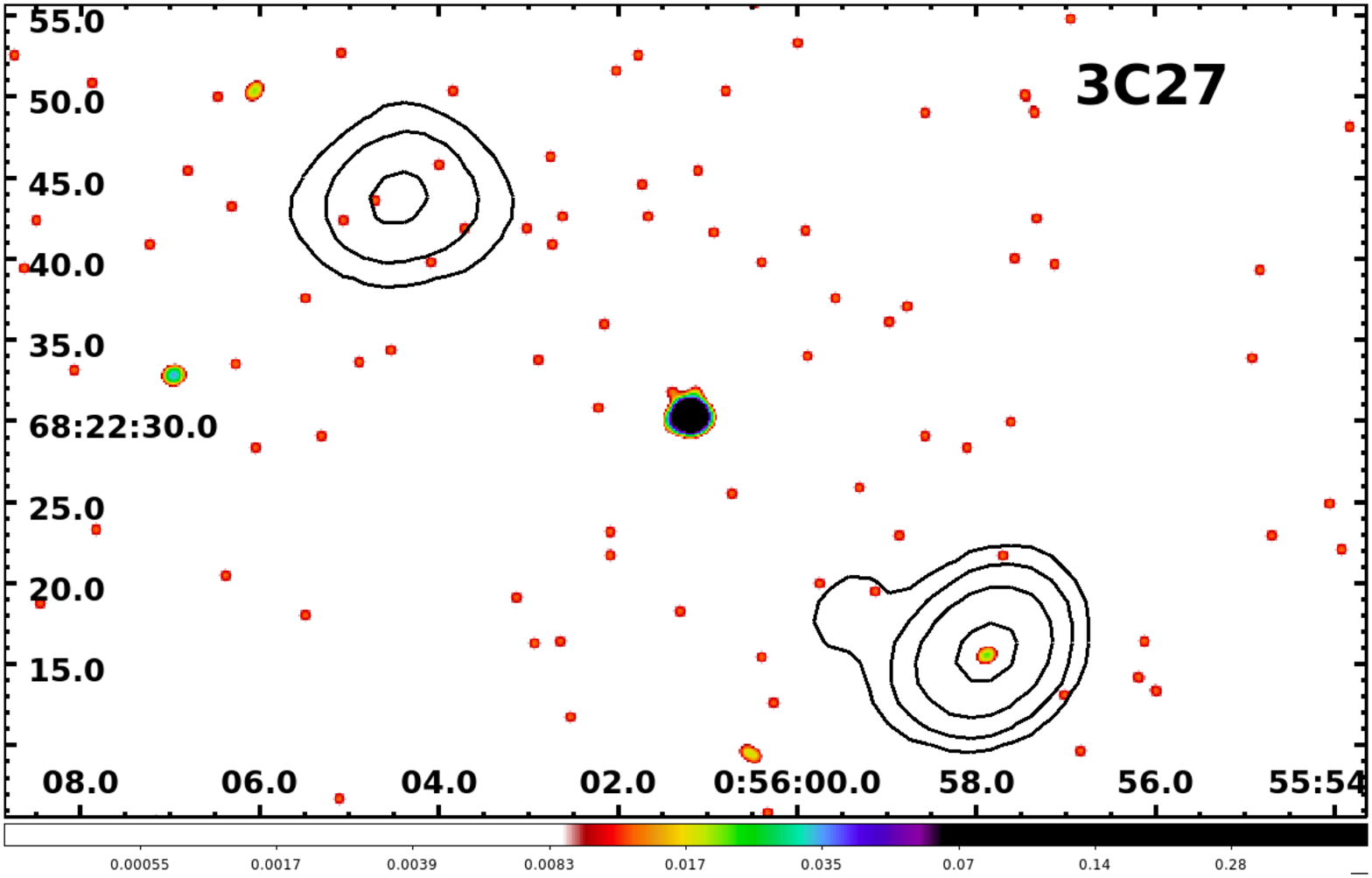}
\caption[3CR\,27]{The X-ray image of 3CR\,27 for the energy band 0.5-7 keV. The image was smoothed with a Gaussian of FWHM=1\asec .0. The radio contours (black) come from a 1.4 GHz VLA map and start at 50 mJy/beam, increasing by factors of four. The image was not registered because of the lack of an obvious radio core but the X-ray nuclear emission of this source was plausibly detected between the two radio lobes.}
\label{fig:3CR27}
\end{figure}

\begin{figure}
\centering
\includegraphics[scale=0.55]{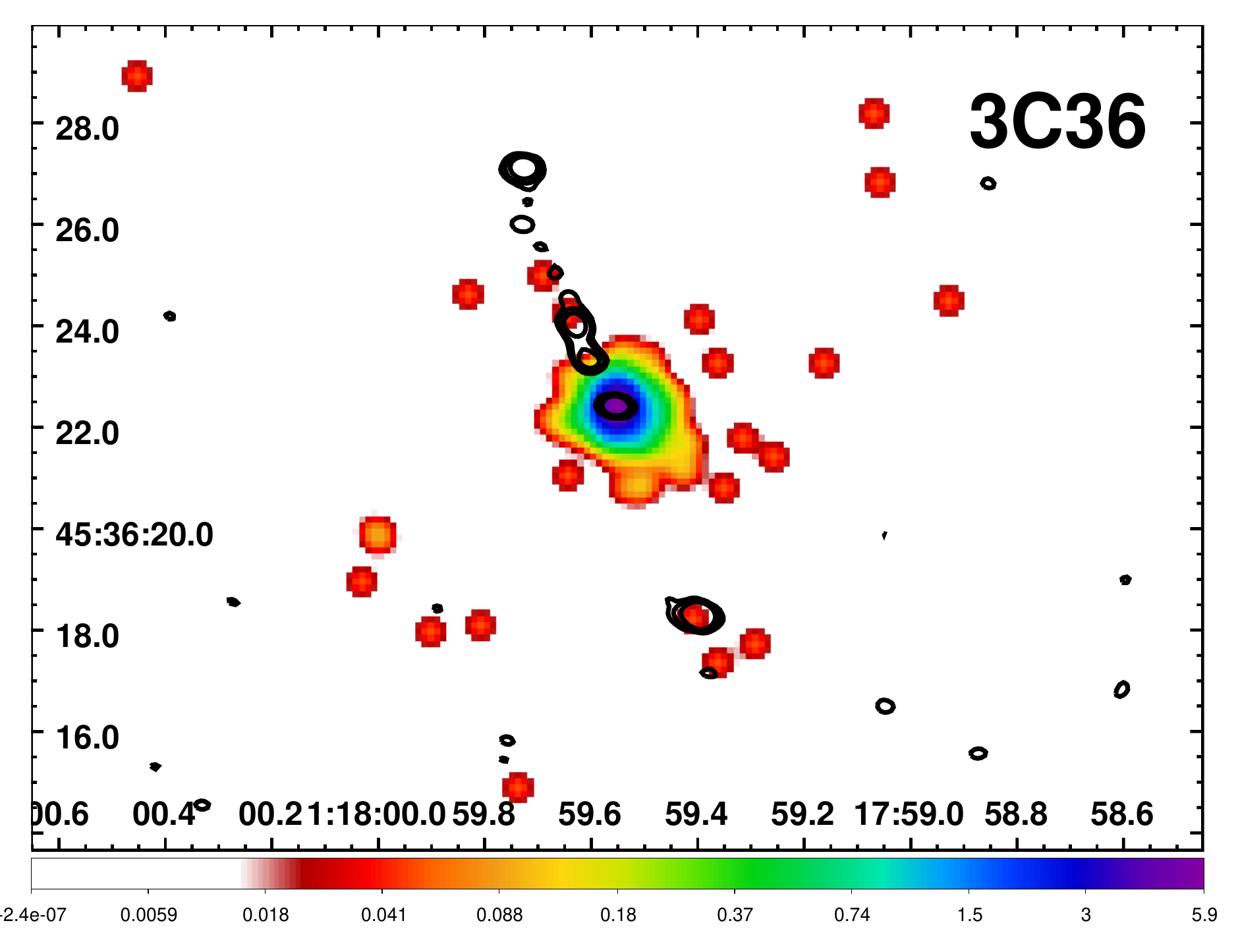}
\caption[3CR\,36]{The X-ray image of 3CR\,36 for the energy band 0.5-7 keV. The image was smoothed with a Gaussian of FWHM=0\asec .5. The radio contours (black) come from an 8.4 GHz VLA map and start at 1 mJy/beam, increasing by factors of four.}
\label{fig:3CR36}
\end{figure}

\begin{figure}
\centering
\includegraphics[scale=0.4, angle=-90]{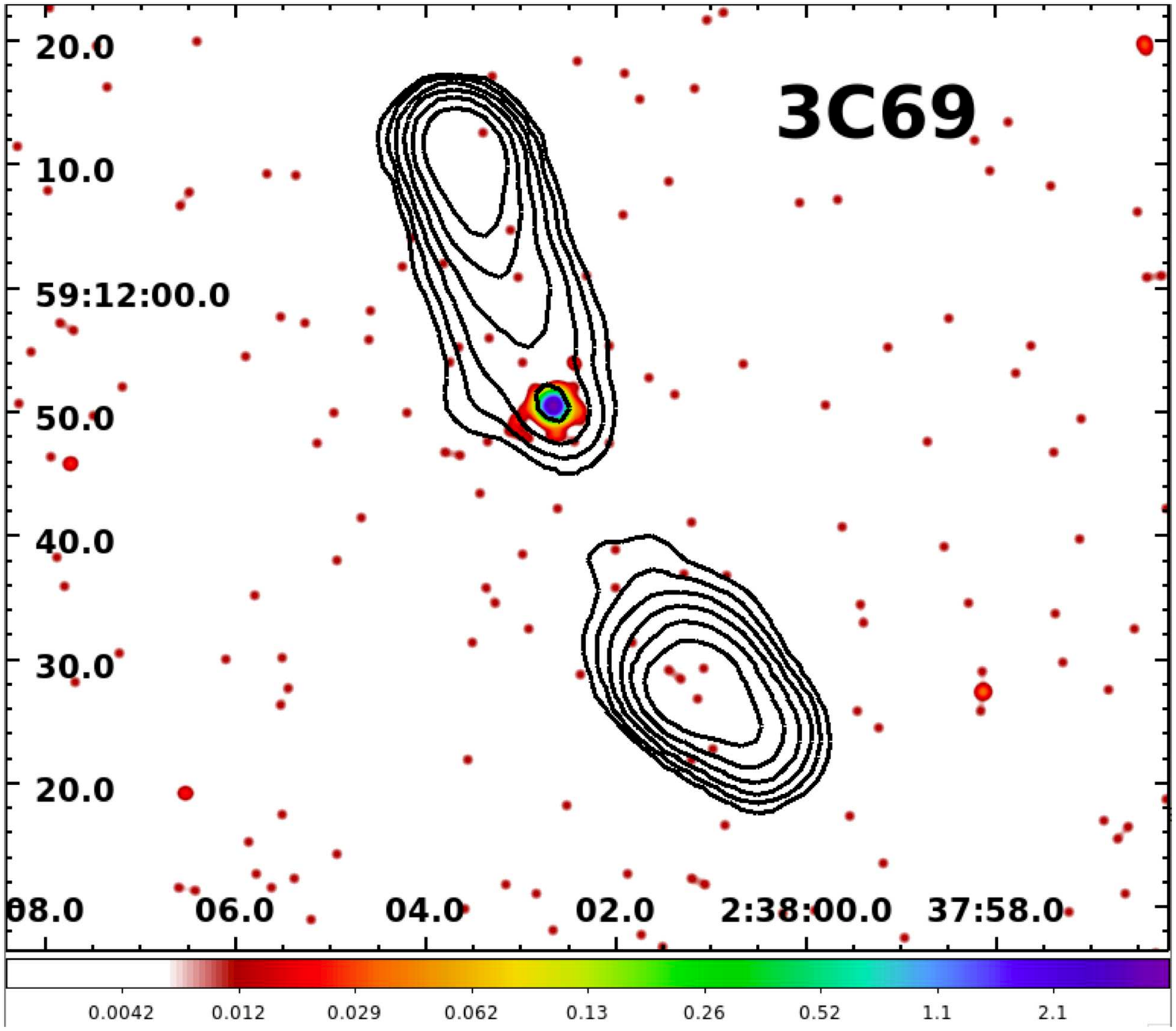}
\caption[3CR\,69]{The X-ray image of 3CR\,69 for the energy band 0.5-7 keV. The \chn\ image was smoothed with a Gaussian of FWHM=1\asec .0. The radio contours (black) come from a 4.9 GHz VLA map and start at 2 mJy/beam, increasing by factors of two.}
\label{fig:3CR69}
\end{figure}

\begin{figure}
\centering
\includegraphics[scale=0.35, angle=0]{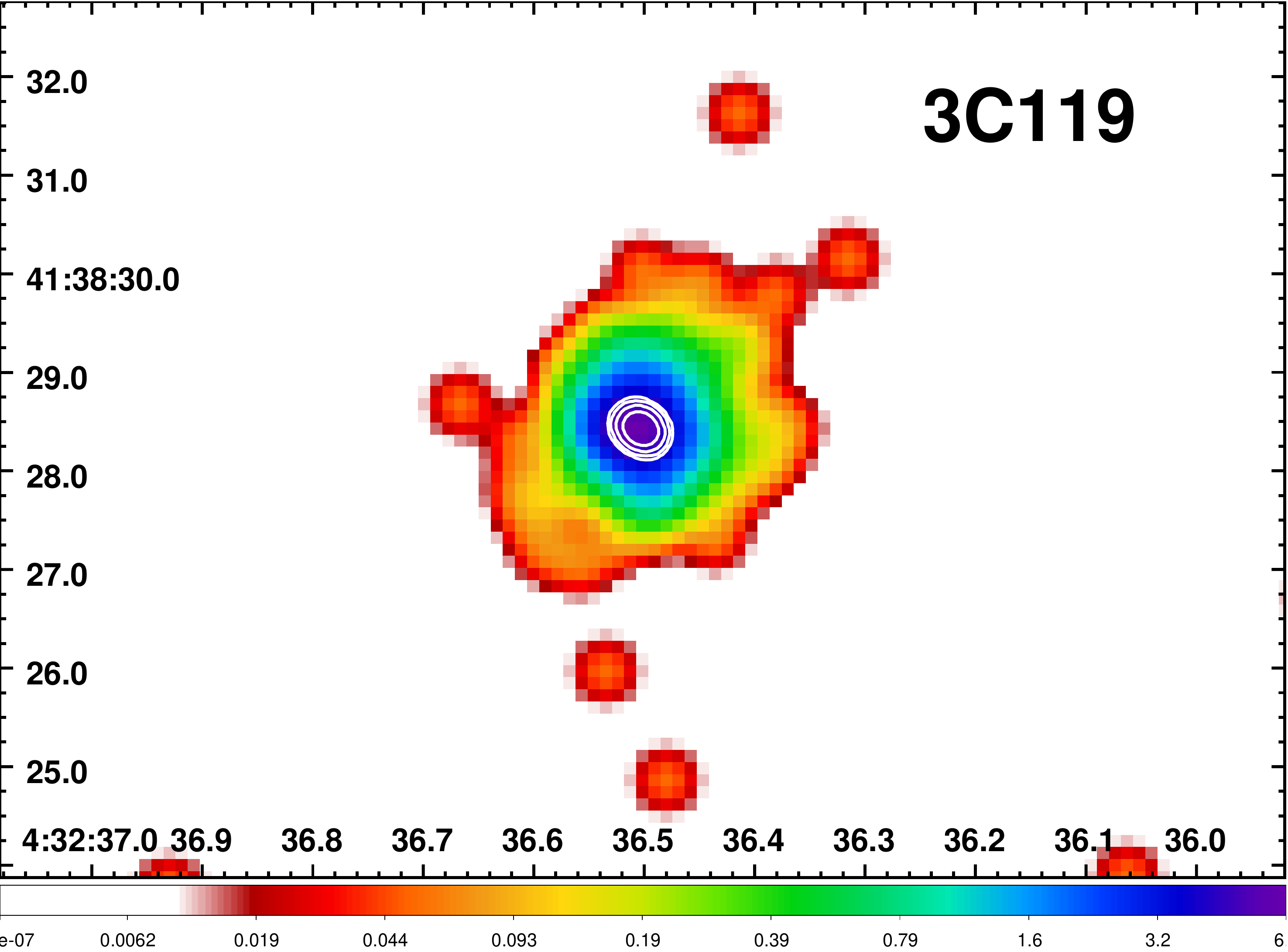}
\caption[3CR\,119]{The X-ray image of 3CR\,119 for the energy band 0.5-7 keV. The image was smoothed with a Gaussian of FWHM=0\asec .5. The radio contours (white) come from an 8.4 GHz VLA map and start at 10 mJy/beam, increasing by factors of four.}
\label{fig:3CR119}
\end{figure}

\begin{figure}
\centering
\includegraphics[scale=0.4, angle=-90]{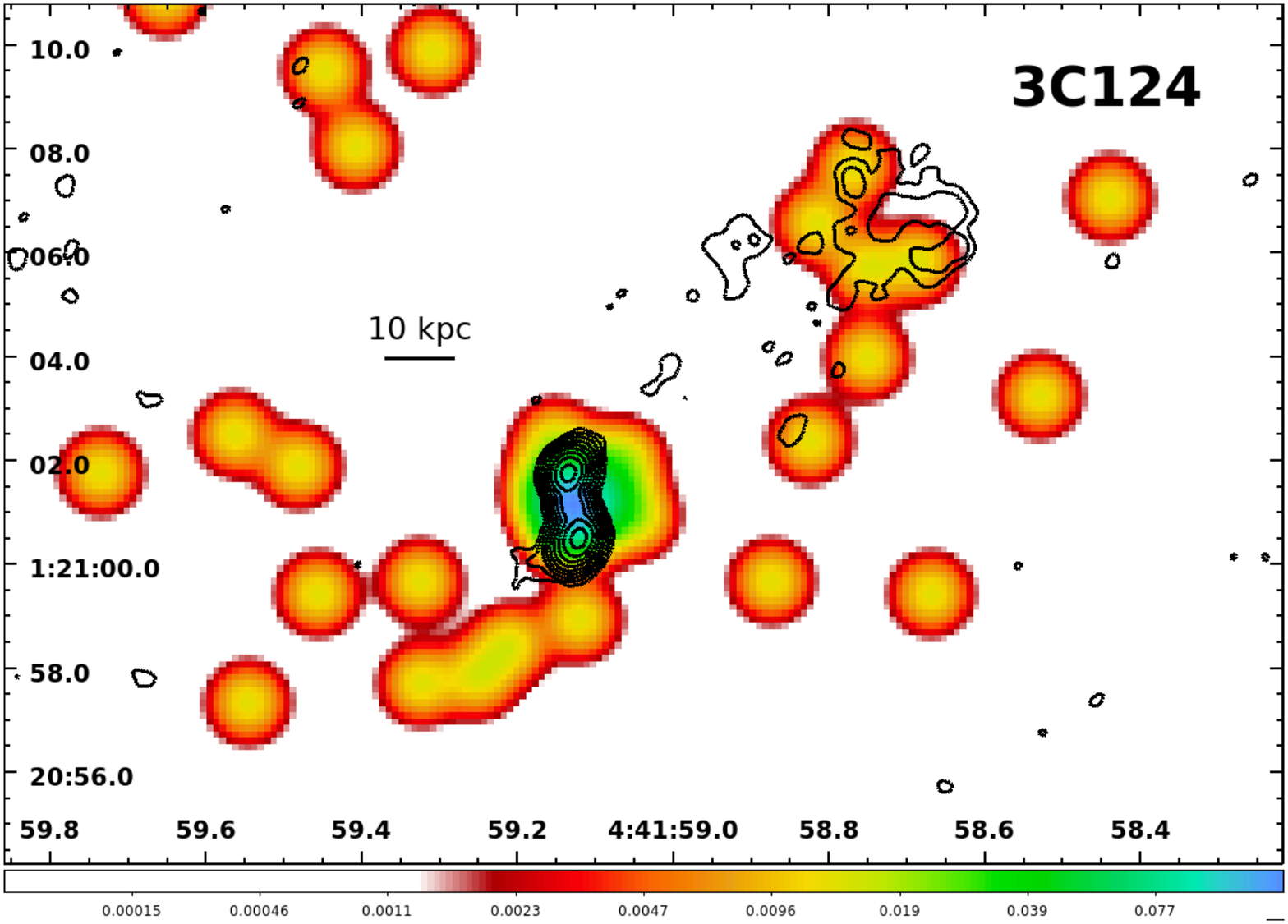}
\caption[3CR\,124]{The X-ray image of 3CR\,124 for the energy band 0.5-7 keV. The image was smoothed with a Gaussian of FWHM=1\asec .0. The radio contours (black) come from a 8.4 GHz map, kindly provided by C. C. Cheung, and start at 0.12 mJy/beam, increasing by factors of two. The NW external lobe is detected with 5 counts in the X-ray and diffuse extended emission is observed on 100 kpc scale in WNW-ESE direction.}
\label{fig:3CR124}
\end{figure}

\begin{figure}
\centering
\includegraphics[scale=0.35, angle=-90]{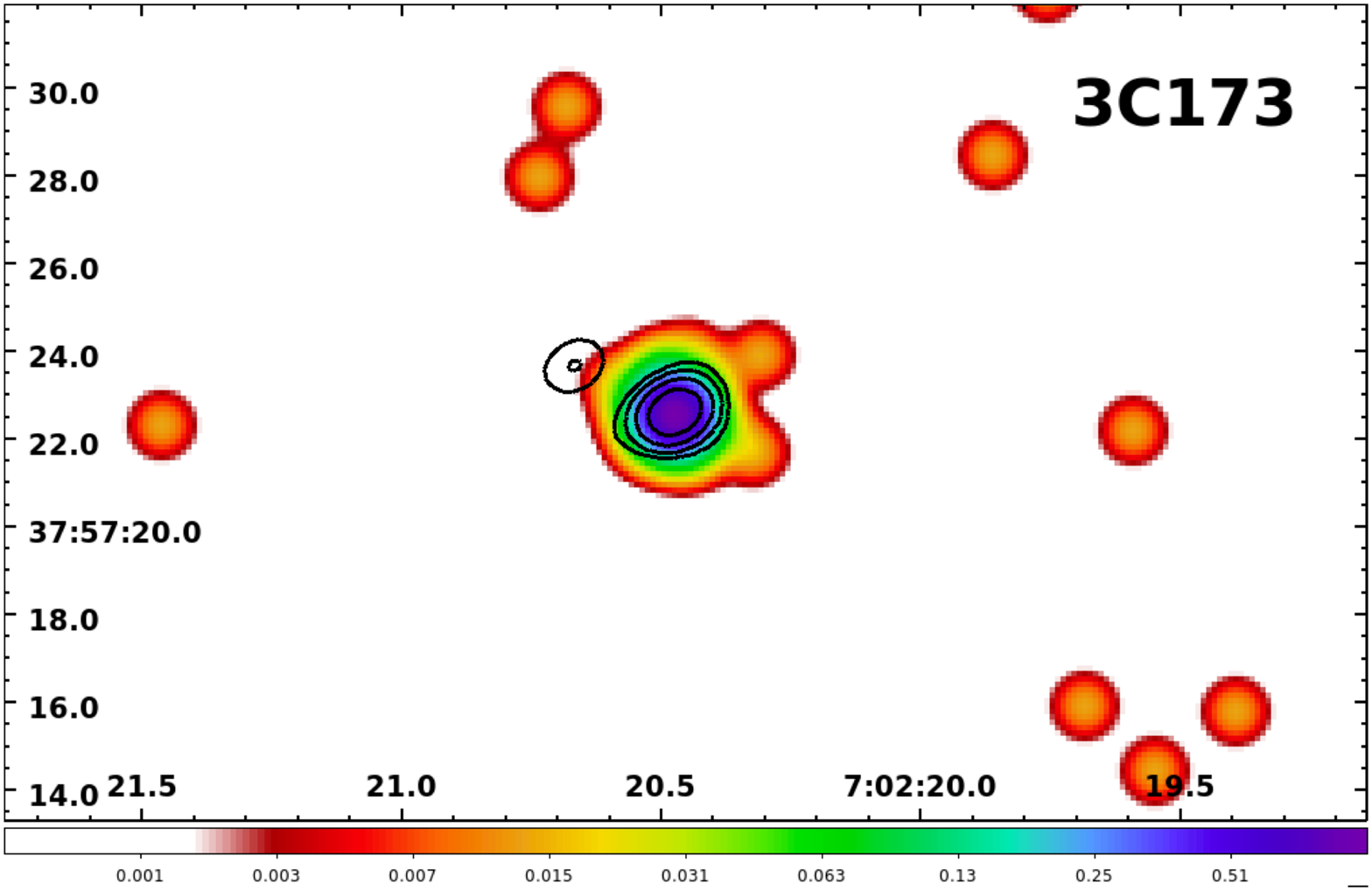}
\caption[3CR\,173]{The X-ray image of 3CR\,173 for the energy band 0.5-7 keV. The image was smoothed with a Gaussian of FWHM=1\asec .0. The radio contours (black) come from a 4.9 GHz VLA map and start at 25 mJy/beam, increasing by factors of two.}
\label{fig:3CR173}
\end{figure}

\begin{figure}
\centering
\includegraphics[scale=0.4, angle=-90]{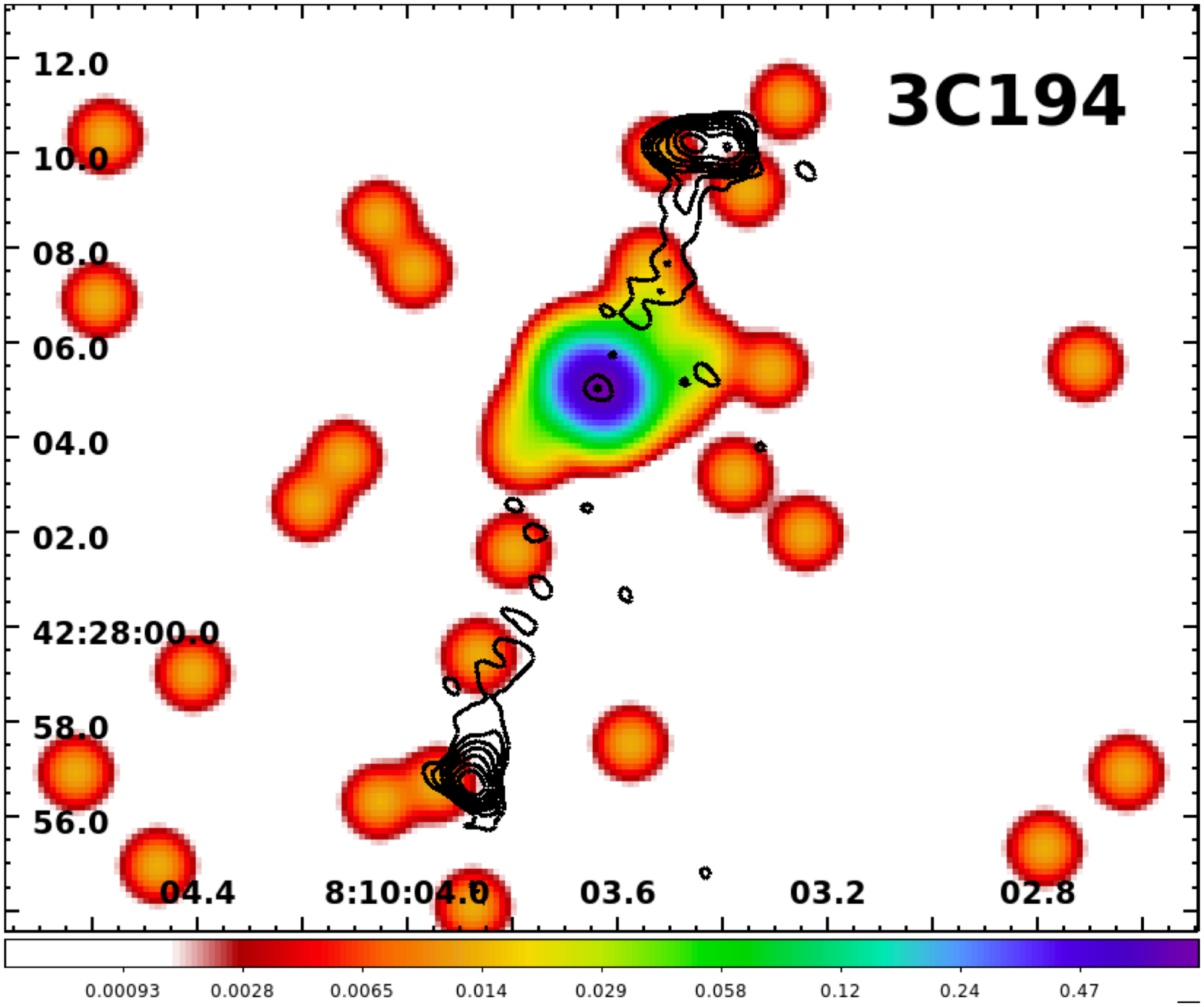}
\caption[3CR\,194]{The X-ray image of 3CR\,194 for the energy band 0.5-7 keV. The image was smoothed with a Gaussian of FWHM=1\asec .0. The radio contours (black) come from a 4.9 GHz VLA map and start at 1.4 mJy/beam, increasing by factors of two.}
\label{fig:3CR194}
\end{figure}

\begin{figure}
\centering
\includegraphics[scale=0.4, angle=-90]{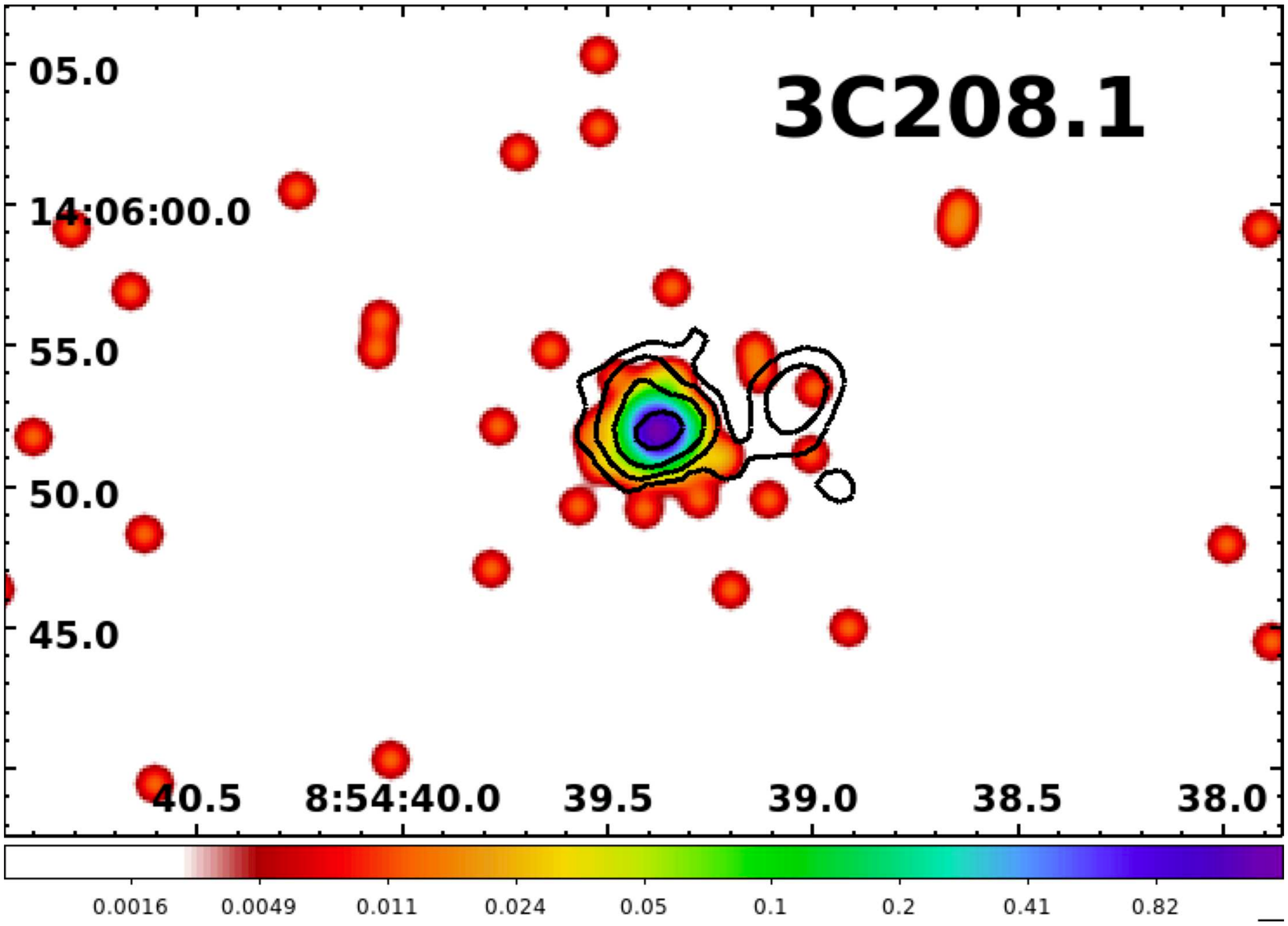}
\caption[3CR\,208.1]{The X-ray image of 3CR\,208.1 for the energy band 0.5-7 keV. The \chn\ image was smoothed with a Gaussian of FWHM=1\asec .0. Radio contours (black) come from a 14.9 GHz VLA map and start at 1.5 mJy/beam, increasing by factors of four.}
\label{fig:3CR208.1}
\end{figure}

\begin{figure}
\centering
\includegraphics[scale=0.4, angle=-90]{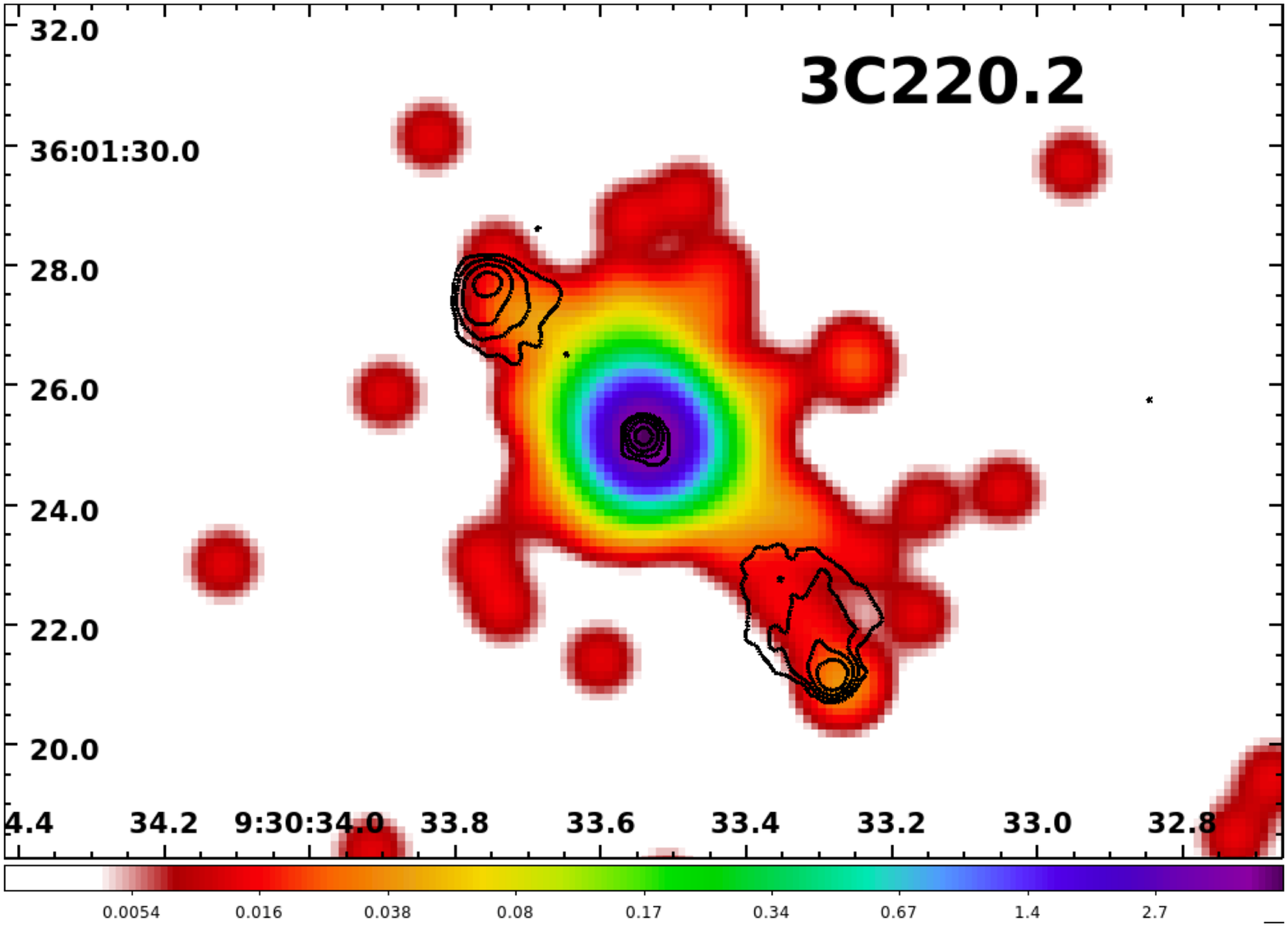}
\caption[3CR\,220.2]{The X-ray image of 3CR\,220.2 for the energy band 0.5-7 keV. The image was smoothed with a Gaussian of FWHM=1\asec .0. The radio contours (black) come from a 8.4 GHz map, kindly supplied by C. C. Cheung, and start at 0.3 mJy/beam, increasing by factors of four. The nucleus and the SW hotspot are detected in the X-ray. An elongation of the X-ray emission is observed along the lobe axis, but, within 3\asec, the source is statistically consistent with the PSF.}
\label{fig:3CR220.2}
\end{figure}

\begin{figure}
\centering
\includegraphics[scale=0.37, angle=-90]{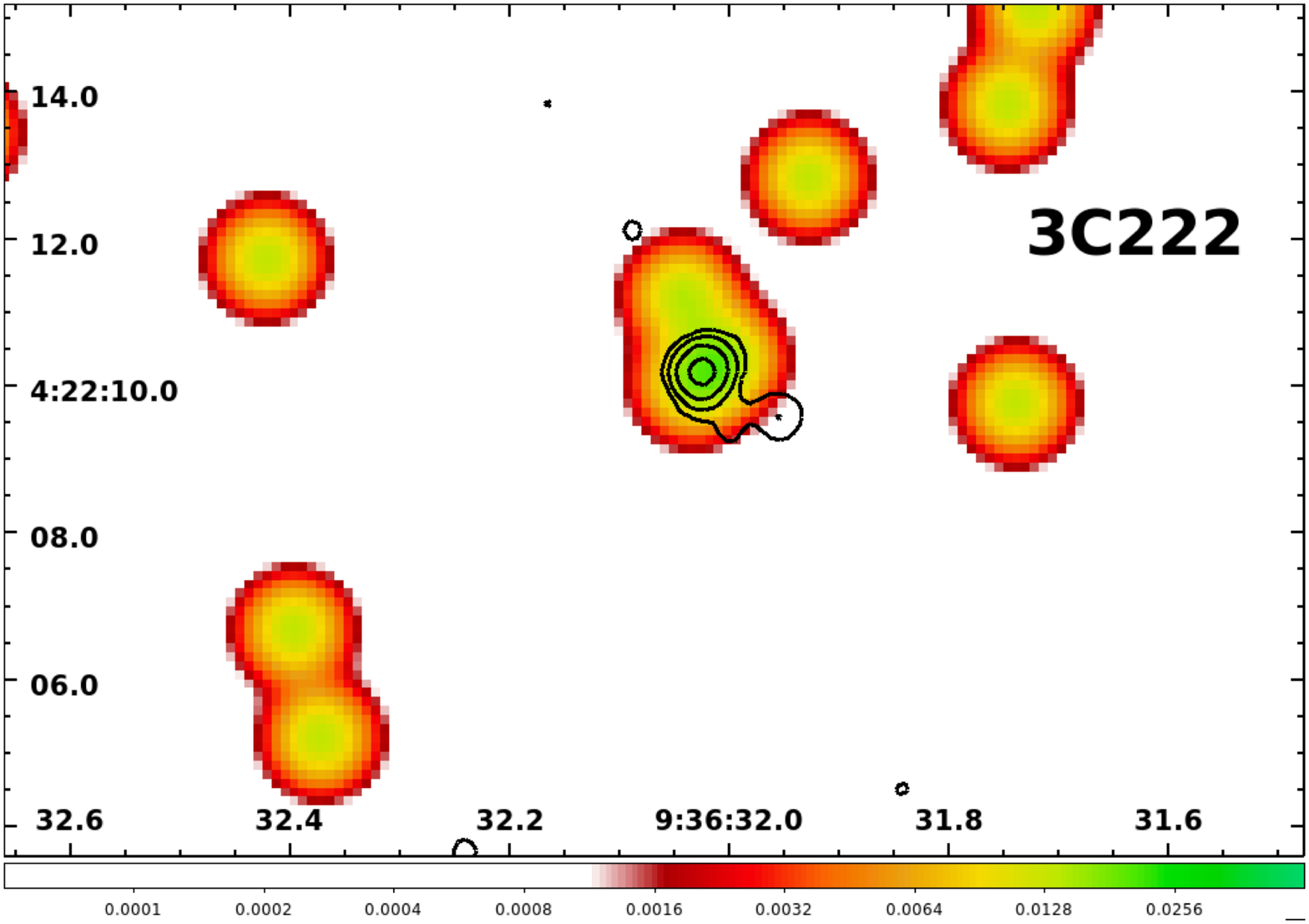}
\caption[3CR\,222]{The X-ray image of 3CR\,222 for the energy band 0.5-7 keV. The \chn\ image was smoothed with a Gaussian of FWHM=1\asec .0. The radio contours (black) come from a 4.9 GHz VLA map and start at 1 mJy/beam, increasing by factors of four.}
\label{fig:3CR222}
\end{figure}

\begin{figure}
\centering
\includegraphics[scale=0.4, angle=0]{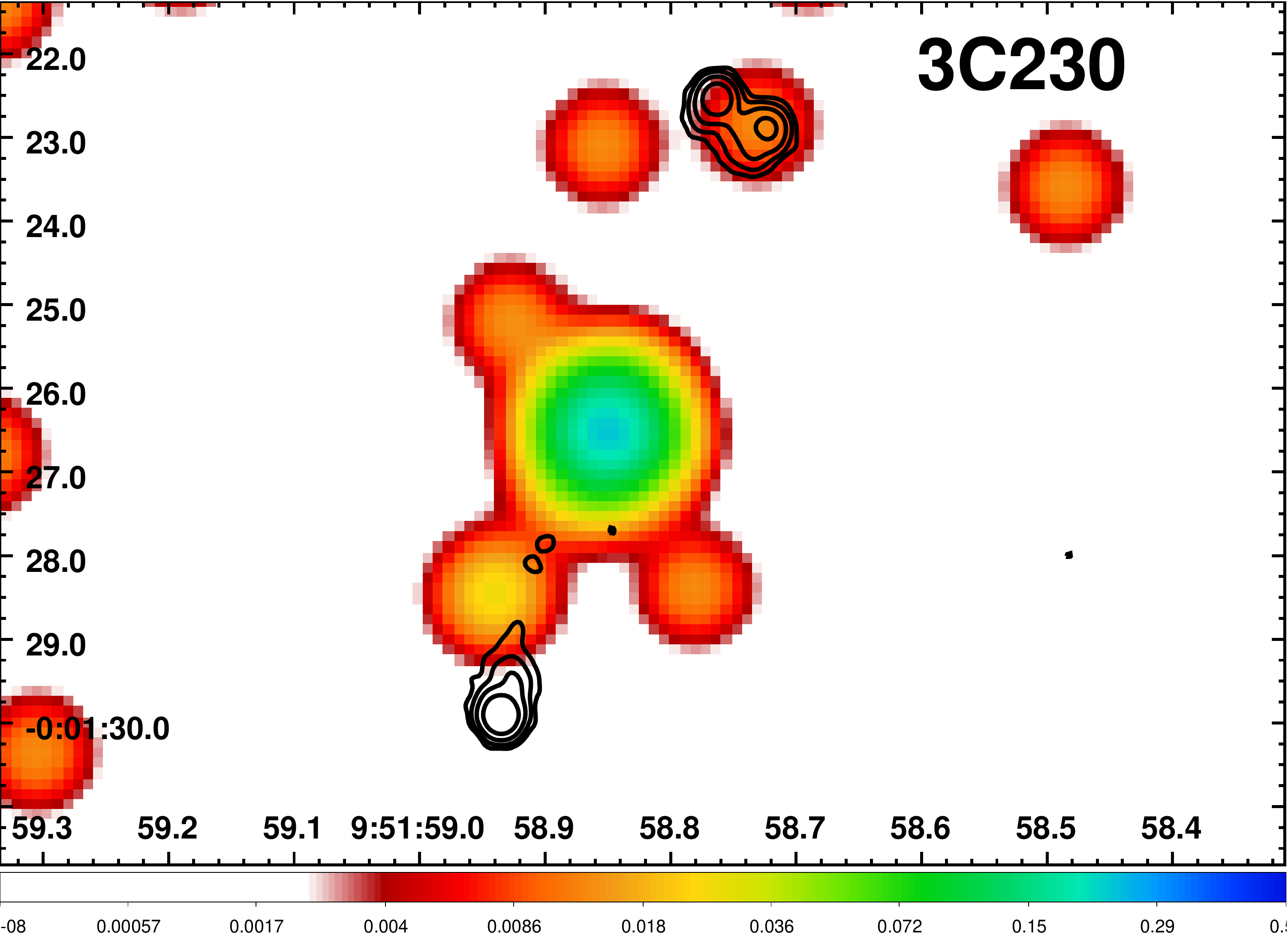}
\caption[3CR\,230]{The X-ray image of 3CR\,230 for the energy band 0.5-7 keV. The image was smoothed with a Gaussian of FWHM=1\asec .0. The radio contours (black) come from an 8.4 GHz map, kindly provided by C. C. Cheung, and start at 0.3 mJy/beam, increasing by factors of four. The X-ray image has not been registered because the radio core position is uncertain.}
\label{fig:3CR230}
\end{figure}

\begin{figure}
\centering
\includegraphics[scale=0.4, angle=0]{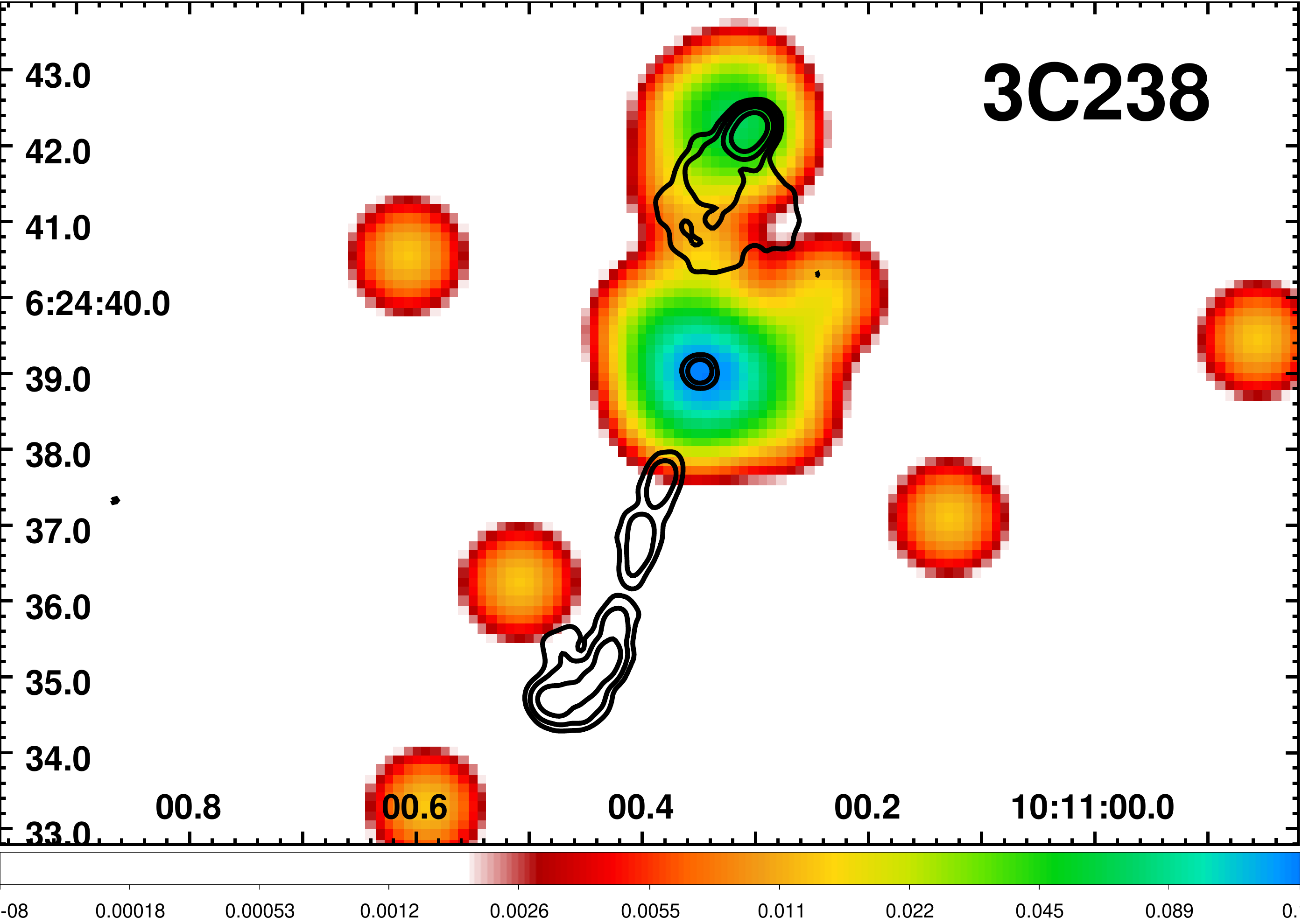}
\caption[3CR\,238]{The X-ray image of 3CR\,238 for the energy band 0.5-7 keV. The image was smoothed with a Gaussian of FWHM=1\asec .0. The radio contours (black) come from an 8.4 GHz map, kindly supplied by C. C. Cheung, and start at 0.3 mJy/beam, increasing by factors of four. The northern hotspot and the nucleus are detected in the X-ray.}
\label{fig:3CR238}
\end{figure}

\begin{figure}
\centering
\includegraphics[scale=0.4, angle=-90]{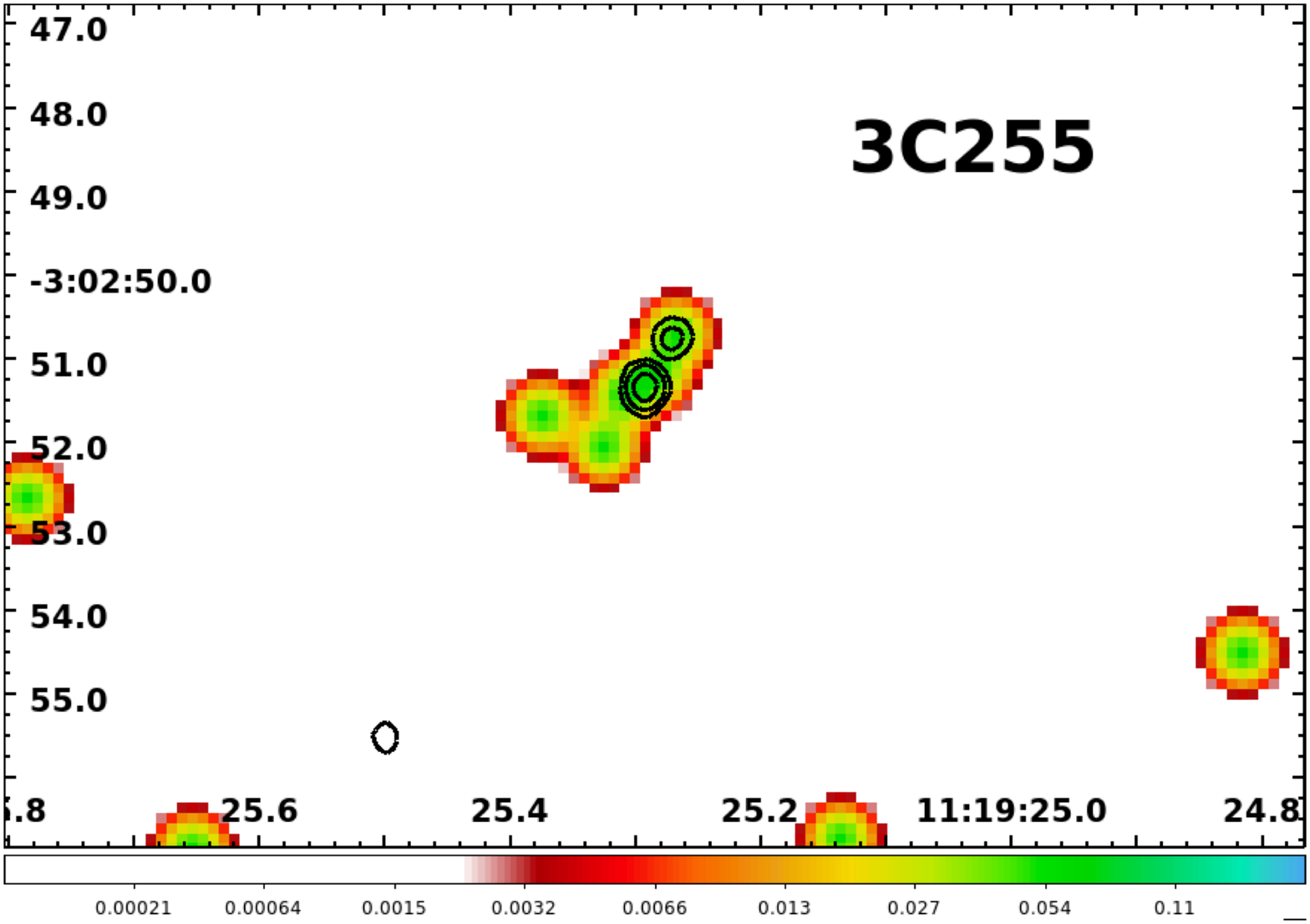}
\caption[3CR\,255]{The X-ray image of 3CR\,255 for the energy band 0.5-7 keV. The image was smoothed with a Gaussian of FWHM=0\asec .5. The radio contours (black) come from a 8.4 GHz VLA map and start at 1 mJy/beam, increasing by factors of four.}
\label{fig:3CR255}
\end{figure}

\begin{figure}
\centering
\includegraphics[scale=0.45, angle=0]{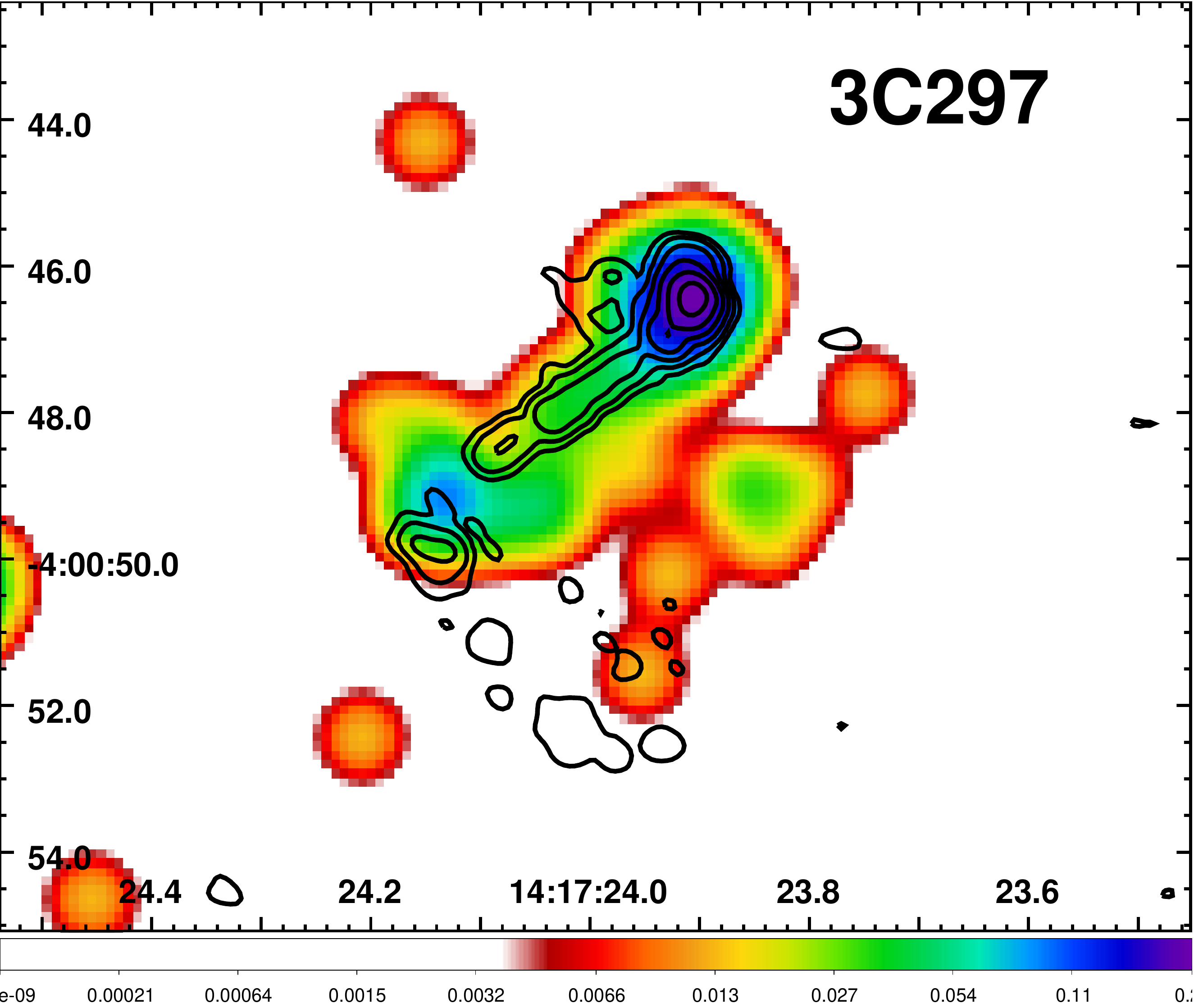}
\caption[3CR\,297]{The X-ray image of 3CR\,297 for the energy band 0.5-7 keV. The image was smoothed with a Gaussian of FWHM=1\asec .0. The radio contours (black) come from a 8.4 GHz VLA map \cite{hilbert16} and start at 0.08 mJy/beam, increasing by factors of four. The image is registered with the NW knot position. The nucleus is detected in the X-ray although its position is uncertain. The X-ray emission in the westward direction has no optical or radio counterpart.}
\label{fig:3CR297}
\end{figure}

\begin{figure}
\centering
\includegraphics[scale=0.35, angle=0]{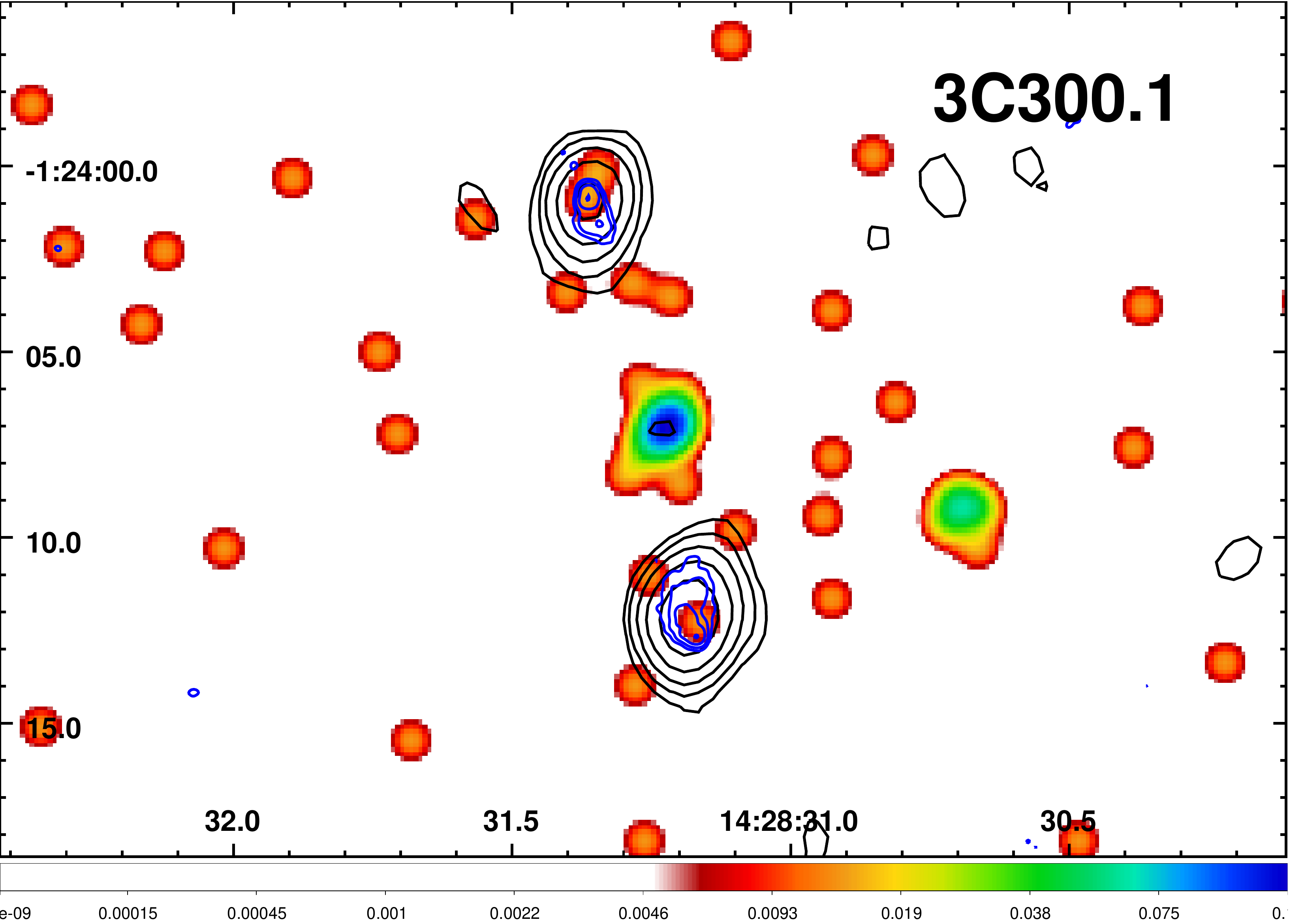}
\caption[3CR\,300.1]{The X-ray image of 3CR\,300.1 for the energy band 0.5-7 keV. The \chn\ image was smoothed with a Gaussian of FWHM=1\asec .0. The radio black contours come from a 14.9 GHz VLA map and start at 3.5 mJy/beam, increasing by factors of two, the blue ones come from an 8.4 GHz map and start at 0.6 mJy/beam, increasing by factors of four. The radio core is visible only in the 14.9 GHz map and it is X-ray detected, while, $\sim$7\asec\ SW from the nucleus, there is a point-like X-ray field source.}
\label{fig:3CR300.1}
\end{figure}

\begin{figure}
\centering
\includegraphics[scale=0.4, angle=0]{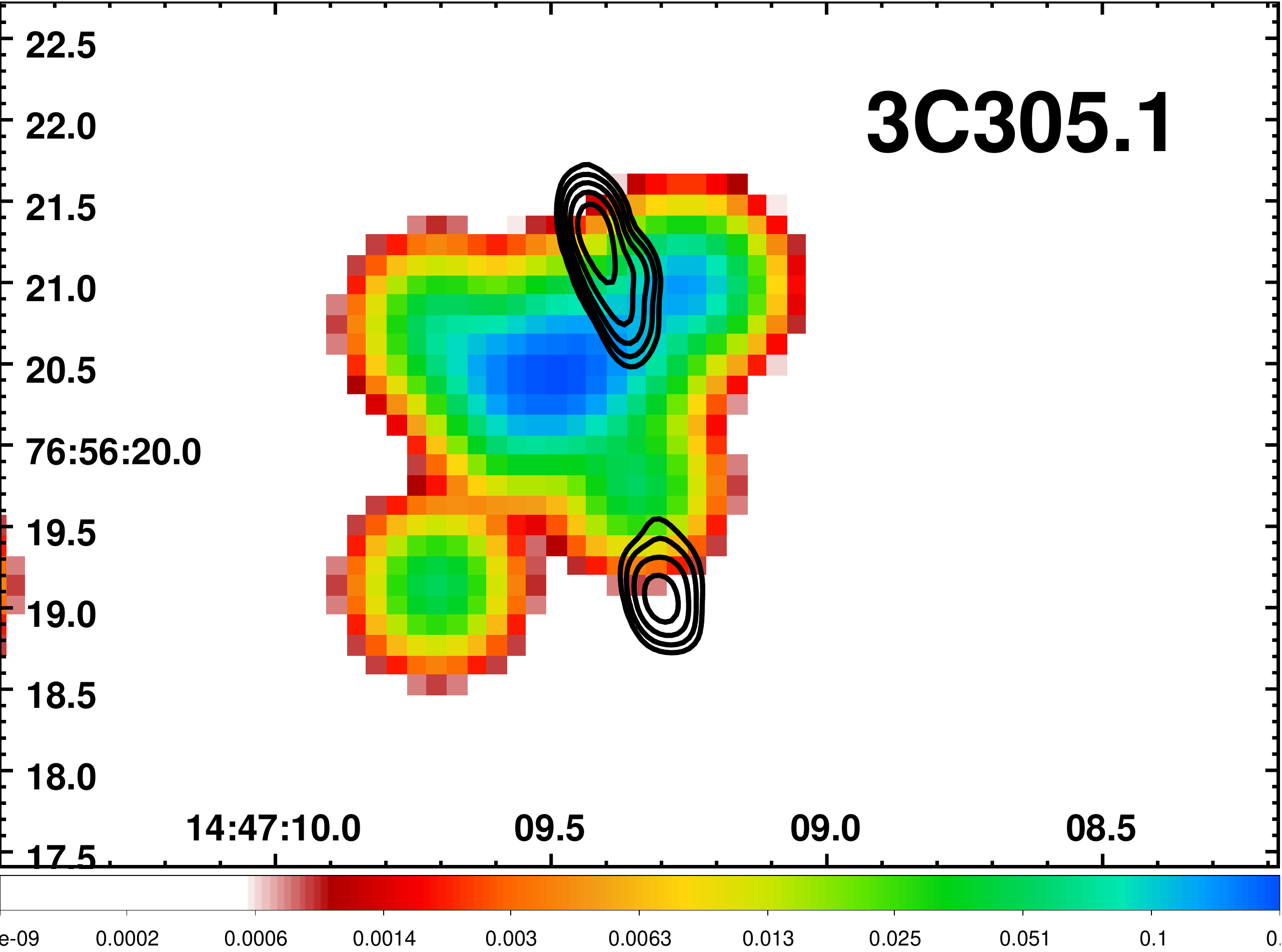}
\caption[3CR\,305.1]{The X-ray image of 3CR\,305.1 for the energy band 0.5-7 keV. The image was smoothed with a Gaussian of FWHM=0\asec .5. The radio contours (black) come from an 8.4 GHz VLA map and start at 2 mJy/beam, increasing by factors of two. No astrometric registration was performed for this source.}
\label{fig:3CR305.1}
\end{figure}

\clearpage

\section{B. X-ray observations of the 3CR sample}
\label{sec:appendixB}

In Table~\ref{tab:3CRtab} we list the summary of the X-ray observations for the 3CR extragalactic sample carried out with \chn\ and {\it XMM-Newton}.

\chn\ X-ray observations have been uniformly analysed, as reported in our previous and current investigations, as part of the \chn\ 3CR survey project. The 3CR extragalactic portion of the catalog includes 298 sources \citep{spinrad85}.  Adopting a step-wise strategy, we requested the observation of small subsamples in each cycle, to minimize the impact on the \chn\ schedule. The full observational coverage of all 3CR sources with $z<$0.3 was achieved during \chn\ Cycles 9 and 12, for a total of 56 sources observed \citep{massaro10,massaro12}. An additional 19 3CR radio galaxies were published in 2013 \citep[0.3$<z<$0.5, \chn\ Cycle 13, in][]{massaro13} and a data paper presenting the most recent observations of 23 sources was accepted in 2017 \citep[0.5$<z<$1.0, Cycle 15, in][]{massaro17}.  We also performed a uniform re-analysis of 140 objects, listed in the \chn\ archive, adopting the same data reduction procedures, from which we excluded 7 sources that have been extensively discussed in the literature and three other sources for PI reasons \citep[][and references therein]{massaro11,massaro15}. Several subsets of the 3CR sample have been also observed by other groups \citep[e.g.][]{balmaverde12,wilkes13}. Here we present an update of the summary also including the 16 targets analysed in the present work (Cycle 17). 

Of the remaining 34 sources unobserved by \chn , 25 are still unidentified, lacking an assigned optical/IR counterpart. We recently observed 21 of these 25 targets with {\it Swift} snapshot observations discovering X-ray counterparts of eleven of them, but even using optical and ultraviolet data available from the instruments on board of the {\it Swift} satellite, we could not discern their optical counterparts \citep{maselli16}.

The {\it XMM-Newton} archive covers to date $\sim$35\% of the entire 3CR extragalactic catalog and all the information provided here is only based on literature search \citep[see e.g.][and references therein for more details]{croston05, evans06, belsole07,croston08,laskar10,shelton11,ineson13,mannering13,ineson17}

We report the 3CR name, the updated value of the redshift $z$ and the radio classification of each source, labeling: radio galaxies (RG), according to the Fanaroff \& Riley criterion \citep{fanaroff74}; quasars (i.e., QSO);
Seyfert galaxies (Sy) and BL Lac objects (BL). We indicate as ``UNID'' those sources lacking a clear optical/IR counterpart. Then we also added a column reporting the radio morphology for the radio galaxies (FR\,I $vs$ FR\,II classes) and indicating those objects that also show the radio structure of: {\it (i)} Compact Steep Spectrum (CSS) or X-shaped (XS) radio sources or {\it (ii)} have been classified in the literature as wide-angle tailed or narrow-angle tailed radio galaxies (WAT and NAT, respectively). We devoted a column to the optical classification of radio galaxies distinguishing between HERG or LERG. We used a ``cluster flag'' to label sources for which there is an associated optical group/cluster reported in the literature and/or those for which an archival X-ray observation confirms the presence of hot gas in the intracluster medium.

For what concerns the X-ray, we verified the presence of \chn\ or {\it XMM-Newton} observations indicating with different symbols if the source was already observed by \chn\ (c) and/or {\it XMM-Newton} (x). We report X-ray detections of radio components ,i.e., jet knot, hotspot or lobe, with the symbols $k$, $h$, and $l$, respectively. The symbol $igm$ was used to indicate the detection of X-ray emitting intergalactic medium. For \chn\ observations we also adopted the symbol $e$ for the sources which show extended X-ray emission around the nucleus highlighted in our analyses by the ``Extent Ratio'' measurements (see \S~\ref{sec:extended}).  For {\it XMM-Newton} observations we only adopted $l$ and $igm$ labels due to the lower angular resolution with respect to \chn\ that does not allow us to see counterparts of jet knots and hotspots in a large fraction of the 3CR sources.

\begin{center} 
\begingroup
\renewcommand\arraystretch{1.0}
\begin{longtable}{rrrrrrrrr}
\caption[]{Summary of the 3CR X-ray observations} \label{tab:3CRtab} \\

\hline
\hline
\noalign{\smallskip}
3CR\tablenotemark{a}  & $z$\tablenotemark{b} & Class\tablenotemark{c} & Radio\tablenotemark{d} & Optical\tablenotemark{e} & Cluster\tablenotemark{f} & X-ray\tablenotemark{g} & \chn\ \tablenotemark{h} & {\it XMM-Newton}\tablenotemark{i} \\
name  &  &  & morph. & class & flag & obs. & detections & detections \\
\hline
\noalign{\smallskip}
\endfirsthead

\multicolumn{9}{c}{{\tablename\ \thetable{} -- continued from previous page}}\\
\hline
\hline
\noalign{\smallskip}
3CR\tablenotemark{a}  & $z$\tablenotemark{b} & Class\tablenotemark{c} & Radio\tablenotemark{d} & Optical\tablenotemark{e} & Cluster\tablenotemark{f} & X-ray\tablenotemark{g} & \chn\ \tablenotemark{h} & {\it XMM-Newton}\tablenotemark{i} \\
name  &  &  & morph. & class & flag & obs. & detections & detections \\
\hline
\noalign{\smallskip}
\endhead

\noalign{\smallskip}
\hline
\endfoot

\noalign{\smallskip}
\hline
\endlastfoot

  2.0 & 1.037 & QSO &  &  &   & c & e & \\
  6.1 & 0.840 & RG & FRII & HERG &   & c & h & \\
  9.0 & 2.020 & QSO & LDQ &  &   & c & k;l & \\
  11.1 & ? & UNID & - & - &   & x &  & \\
  13.0 & 1.351 & RG & FRII & HERG &   & c-x & e;h & \\
  14.0 & 1.469 & QSO &  &  &   & c &  & \\
  14.1 & ? & UNID & - & - &   & x &  & \\
  15.0 & 0.074 & RG & FRI & LERG &   & c & k;l & \\
  16.0 & 0.405 & RG & FRII & HERG &   & c-x & e;h;l & \\
  17.0 & 0.220 & QSO &  &  &   & c & k & \\
  18.0 & 0.188 & RG & FRII & HERG &   & c & e & \\
  19.0 & 0.482 & RG & FRII & LERG & \checkmark & c-x & h;igm & \\
  20.0 & 0.174 & RG & FRII & HERG &   & c-x &  & \\
  21.1 & ? & UNID & - & - &   & x &  & \\
  22.0 & 0.936 & RG & FRII &  &   & c &  & \\
  27.0 & 0.184 & RG & FRII & HERG &   & c &  & \\
  28.0 & 0.195 & RG & FRI & LERG & \checkmark & c-x & igm & igm\\
  29.0 & 0.045 & RG & FRI & LERG & \checkmark & c & k;igm & \\
  31.0 & 0.017 & RG & FRI & LERG & \checkmark & c-x & k & igm\\
  33.0 & 0.060 & RG & FRII & HERG & \checkmark & c-x & h:l & l;igm\\
  33.1 & 0.181 & RG & FRII & HERG &   & c &  & \\
  33.2 & ? & UNID & - & - &   & x &  & \\
  34.0 & 0.69 & RG & FRII & HERG & \checkmark & c & igm & \\
  35.0 & 0.067 & RG & FRII & LERG & \checkmark & c-x & e;l & l;igm\\
  36.0 & 1.301 & RG & FRII & HERG &   & c &  & \\
  40.0 & 0.018 & RG & FRI-WAT & LERG & \checkmark & c-x & igm & igm\\
  41.0 & 0.795 & RG & FRII & HERG &   & c & h & \\
  42.0 & 0.396 & RG & FRII & HERG &   & c-x &  & \\
  43.0 & 1.459 & QSO & CSS &  &   & c &  & \\
  44.0 & 0.66 & QSO &  &  & \checkmark & c &  & \\
  46.0 & 0.437 & RG & FRII & HERG & \checkmark & c-x &  & igm\\
  47.0 & 0.425 & QSO & LDQ &  &   & c & h;l & \\
  48.0 & 0.367 & QSO & CSS &  &   & c &  & \\
  49.0 & 0.236 & RG & FRII-CSS & HERG &   & c &  & \\
  52.0 & 0.29 & RG & FRII-XS & HERG & \checkmark & c & h & \\
  54.0 & 0.827 & RG & FRII & HERG &   & c & h & \\
  55.0 & 0.735 & RG & FRII & HERG &   & c-x &  & \\
  61.1 & 0.188 & RG & FRII & HERG &   & c-x & h & \\
  63.0 & 0.175 & RG & FRII-XS & HERG &   & c &  & \\
  65.0 & 1.176 & RG & FRII & HERG &   & c-x & e;h & \\
  66A & ? & BL &  & - & \checkmark & c-x &  & \\
  66B & 0.0213 & RG & FRI & LERG & \checkmark & c-x & k & igm\\
  67.0 & 0.310 & RG & FRII-CSS &  &   & c-x &  & \\
  68.1 & 1.238 & QSO &  &  &   & c &  & \\
  68.2 & 1.575 & RG & FRII & HERG &   & c & e;h & \\
  69.0 & 0.458 & RG & FRII & HERG &   & c &  & \\
  71.0 & 0.004 & SEY & - & Sy2 &   & c-x &  & \\
  75.0 & 0.023 & RG & FRI-WAT & LERG & \checkmark & c-x & igm & igm\\
  76.1 & 0.032 & RG & FRI & LERG & \checkmark & c-x & igm & igm\\
  78.0 & 0.029 & RG & FRI & LERG &   & c & k & \\
  79.0 & 0.256 & RG & FRII & HERG & \checkmark & c-x &  & igm\\
  83.1 & 0.025 & RG & FRI-NAT & LERG & \checkmark & c-x & k;igm & igm\\
  84.0 & 0.018 & RG & FRI & LERG & \checkmark & c-x & igm & igm\\
  86.0 & ? & UNID & - & - &   &  &  & \\
  88.0 & 0.030 & RG & FRI & LERG & \checkmark & c-x & k;igm & igm\\
  89.0 & 0.140 & RG & FRI-WAT & LERG & \checkmark & c & igm & \\
  91.0 & ? & UNID & - & - &   &   &  & \\
  93.0 & 0.357 & QSO &  &  &   & c & e & \\
  93.1 & 0.243 & RG & FRII-CSS & HERG & \checkmark & c &  & \\
  98.0 & 0.030 & RG & FRII & HERG &   & c-x & l & l\\
  99.0 & 0.426 & SEY & - & Sy2 &   & c &  & \\
  103.0 & 0.33 & RG & FRII &  &   & c &  & \\
  105.0 & 0.089 & RG & FRII & HERG &   & c-x & k;h & \\
  107.0 & 0.785 & RG & FRII & HERG &   & c & e & \\
  109.0 & 0.307 & RG & FRII & HERG &   & c-x & h;l & \\
  111.0 & 0.049 & RG & FRII &  &   & c-x & k;h & \\
  114.0 & 0.815 & RG & FRII & LERG &   & c &  & \\
  119.0 & 1.023 & QSO & CSS &  &   & c &  & \\
  123.0 & 0.218 & RG & FRII & LERG & \checkmark & c & h;igm & \\
  124.0 & 1.083 & RG & FRII & HERG &   & c & l & \\
  125.0 & ? & UNID & - & - &   &   &  & \\
  129.0 & 0.021 & RG & FRI-NAT &  & \checkmark & c-x & k;igm & igm\\
  129.1 & 0.022 & RG & FRI &  & \checkmark & c-x & igm & igm\\
  130.0 & 0.109 & RG & FRI-WAT &  & \checkmark & c & e & \\
  131.0 & ? & UNID & - & - &   &   &  & \\
  132.0 & 0.214 & RG & FRII & LERG & \checkmark & c-x &  & \\
  133.0 & 0.278 & RG & FRII & HERG &   & c &  & \\
  134.0 & ? & UNID & - & - &   &   &  & \\
  135.0 & 0.127 & RG & FRII & HERG & \checkmark & c &  & \\
  136.1 & 0.064 & RG & FRII-XS & HERG &   & c & e & \\
  137.0 & ? & UNID & - & - &   &   &  & \\
  138.0 & 0.759 & QSO & CSS &  &   & c &  & \\
  139.2 & ? & UNID & - & - &   &   &  & \\
  141.0 & ? & UNID & - & - &   &   &  & \\
  142.1 & 0.406 & RG & FRII &  &   & c &  & \\
  147.0 & 0.545 & QSO & CSS &  &   & c &  & \\
  152.0 & ? & UNID & - & - &   &   &  & \\
  153.0 & 0.277 & RG & FRII & LERG & \checkmark & c-x &  & \\
  154.0 & 0.58 & QSO &  &  &   & c & e;k & \\
  158.0 & ? & UNID & - & - &   &   &  & \\
  165.0 & 0.296 & RG & FRII & LERG &   & c & e & \\
  166.0 & 0.245 & RG & FRII & LERG &   & c &  & \\
  169.1 & 0.633 & RG & FRII & HERG &   & c &  & \\
  171.0 & 0.238 & RG & FRII & HERG &   & c-x & e & \\
  172.0 & 0.519 & RG & FRII & HERG &   & c &  & \\
  173.0 & 1.035 & QSO & CSS & HERG &   & c &  & \\
  173.1 & 0.292 & RG & FRII & LERG & \checkmark & c & h;l & \\
  175.0 & 0.77 & QSO &  &  &   & c &  & \\
  175.1 & 0.92 & RG & FRII & HERG &   & c &  & \\
  180.0 & 0.22 & RG & FRII & HERG &   & c &  & \\
  181.0 & 1.382 & QSO &  &  &   & c & h & \\
  184.0 & 0.994 & RG & FRII & HERG & \checkmark & c-x & l & igm\\
  184.1 & 0.118 & RG & FRII & HERG & \checkmark & c &  & \\
  186.0 & 1.066 & QSO & CSS &  & \checkmark & c & igm & \\
  187.0 & 0.465 & RG & FRII & LERG &   & c & e;l & \\
  190.0 & 0.246 & QSO & CSS &  &   & c &  & \\
  191.0 & 1.968 & QSO &  &  &   & c & k;l & \\
  192.0 & 0.060 & RG & FRII-XS & HERG & \checkmark & c-x & l & igm\\
  194.0 & 1.184 & RG & FRII & HERG &   & c &  & \\
  196.0 & 0.871 & QSO &  &  &   & c &  & \\
  196.1 & 0.198 & RG & FRII & LERG & \checkmark & c & igm & \\
  197.1 & 0.128 & RG & FRII & HERG & \checkmark & c &  & \\
  198.0 & 0.081 & RG & FRII & HERG & \checkmark & c &  & \\
  200.0 & 0.458 & RG & FRII & LERG & \checkmark & c & k;l & \\
  204.0 & 1.112 & QSO &  &  &   & c-x &  & \\
  205.0 & 1.532 & QSO &  &  &   & c-x &  & \\
  207.0 & 0.680 & QSO & LDQ &  & \checkmark & c-x & k;l & igm\\
  208.0 & 1.112 & QSO &  &  &   & c &  & \\
  208.1 & 1.02 & QSO &  &  &   & c-x &  & \\
  210.0 & 1.169 & RG & FRII & HERG & \checkmark & c-x & e;h & igm\\
  212.0 & 1.049 & QSO & LDQ &  &   & c & e;h & \\
  213.1 & 0.194 & RG & FRI-CSS & LERG & \checkmark & c & e;h & \\
  215.0 & 0.411 & QSO &  &  & \checkmark & c-x & k;l & \\
  216.0 & 0.670 & QSO &  &  &   & c &  & \\
  217.0 & 0.898 & RG & FRII & HERG &   & c &  & \\
  219.0 & 0.175 & RG & FRII & HERG & \checkmark & c & k;l & \\
  220.1 & 0.61 & RG & FRII & HERG & \checkmark & c & igm & \\
  220.2 & 1.156 & QSO &  &  &   & c & h & \\
  220.3 & 0.68 & RG & FRII & HERG &   & c &  & \\
  222.0 & 1.339 & RG & FRI &  &   & c &  & \\
  223.0 & 0.137 & RG & FRII & HERG & \checkmark & c-x &  & igm\\
  223.1 & 0.108 & RG & FRII-XS & HERG &   & c &  & \\
  225A & 1.565 & RG & FRII & HERG &   & c &  & \\
  225B & 0.58 & RG & FRII & HERG &   & c & h & \\
  226.0 & 0.818 & RG & FRII & HERG &   & c &  & \\
  227.0 & 0.086 & RG & FRII & HERG &   & c & h & \\
  228.0 & 0.552 & RG & FRII & HERG &   & c & e;h & \\
  230.0 & 1.487 & RG & FRII & HERG &   & c &  & \\
  231.0 & 0.001 & RG & FRI & LERG &   & c-x &  & \\
  234.0 & 0.185 & RG & FRII & HERG &   & c-x & h & \\
  236.0 & 0.101 & RG & FRII & LERG &   & c &  & \\
  237.0 & 0.877 & RG & FRII-CSS &  &   & c &  & \\
  238.0 & 1.405 & RG & FRII & HERG &   & c & h & \\
  239.0 & 1.781 & RG & FRII & HERG &   & x &  & \\
  241.0 & 1.617 & RG & FRII &  &   & c-x &  & \\
  244.1 & 0.428 & RG & FRII & HERG & \checkmark & c-x & e & \\
  245.0 & 1.028 & QSO &  &  &   & c & k & \\
  247.0 & 0.749 & RG & FRII & HERG & \checkmark & c &  & \\
  249.0 & 1.554 & QSO &  &  &   & x &  & \\
  249.1 & 0.312 & QSO &  &  &   & c-x &  & \\
  250.0 & ? & UNID & - & - &   &   &  & \\
  252.0 & 1.1 & RG & FRII & HERG &   & c &  & \\
  254.0 & 0.737 & QSO & LDQ &  &   & c & e;h & \\
  255.0 & 1.355 & RG & FRII(?) & HERG &   & c &  & \\
  256.0 & 1.819 & RG & FRII & HERG &   & c &  & \\
  257.0 & 2.474 & QSO &  &  &   & x &  & \\
  258.0 & 0.165 & RG & FRI-CSS & LERG & \checkmark & c &  & \\
  263.0 & 0.646 & QSO & LDQ &  &   & c & h & \\
  263.1 & 0.824 & RG & FRII & HERG &   & c &  & \\
  264.0 & 0.022 & RG & FRI & LERG & \checkmark & c-x & k & igm\\
  265.0 & 0.811 & RG & FRII & HERG &   & c & h;l & \\
  266.0 & 1.275 & RG & FRII & HERG &   & c-x &  & \\
  267.0 & 1.14 & RG & FRII & HERG &   & c &  & \\
  268.1 & 0.97 & RG & FRII & HERG &   & c & h & \\
  268.2 & 0.362 & RG & FRII & HERG & \checkmark & c-x & e;h & \\
  268.3 & 0.372 & RG & FRII-CSS &  &   & c &  & \\
  268.4 & 1.402 & QSO &  &  &   & c-x &  & \\
  270.0 & 0.007 & RG & FRI & LERG & \checkmark & c-x & k & igm\\
  270.1 & 1.528 & QSO &  &  &   & c &  & \\
  272.0 & 0.944 & RG & FRII & HERG &   & c &  & \\
  272.1 & 0.003 & RG & FRI & LERG & \checkmark & c-x & k & \\
  273.0 & 0.158 & QSO & CDQ &  &   & c-x & k & \\
  274.0 & 0.004 & RG & FRI & LERG & \checkmark & c-x & k;igm & igm\\
  274.1 & 0.422 & RG & FRII & HERG &   & c-x & e & l\\
  275.0 & 0.48 & RG & FRII & LERG & \checkmark & c &  & \\
  275.1 & 0.555 & QSO & LDQ &  &   & c & k;h;l & \\
  277.0 & 0.414 & RG & FRII &  &   & c &  & \\
  277.1 & 0.320 & QSO & CSS &  &   & c &  & \\
  277.2 & 0.766 & RG & FRII & HERG &   & c-x &  & \\
  277.3 & 0.085 & RG & FRII & HERG &   & c &  & \\
  280.0 & 0.996 & RG & FRII & HERG & \checkmark & c-x & k;h;l & \\
  280.1 & 1.667 & QSO &  &  &   &   &  & \\
  284.0 & 0.240 & RG & FRII & HERG & \checkmark & c-x &  & igm\\
  285.0 & 0.080 & RG & FRII & HERG &   & c & l & \\
  286.0 & 0.850 & QSO & CSS &  &   & c &  & \\
  287.0 & 1.055 & QSO & CSS &  &   & c-x &  & \\
  287.1 & 0.216 & RG & FRII & HERG &   & c & h & \\
  288.0 & 0.246 & RG & FRI & LERG & \checkmark & c & igm & \\
  288.1 & 0.963 & QSO &  &  &   & c &  & \\
  289.0 & 0.967 & RG & FRII & HERG &   & c &  & \\
  292.0 & 0.71 & RG & FRII & HERG & \checkmark & c-x & e & igm\\
  293.0 & 0.045 & RG & FRI & LERG &   & c & e & \\
  293.1 & 0.709 & RG & FRII &  &   & c &  & \\
  294.0 & 1.779 & RG & FRII & HERG & \checkmark & c & h;igm & \\
  295.0 & 0.464 & RG & FRII & LERG & \checkmark & c & h;igm & \\
  296.0 & 0.025 & RG & FRI & LERG & \checkmark & c-x & k;igm & igm\\
  297.0 & 1.406 & QSO &  &  &   & c & k;e & \\
  298.0 & 1.438 & QSO & CSS &  & \checkmark & c-x &  & igm\\
  299.0 & 0.367 & RG & FRII &  & \checkmark & c & h & \\
  300.0 & 0.27 & RG & FRII & HERG &   & c-x &  & \\
  300.1 & 1.159 & RG & FRII & HERG &   & c & & \\
  303.0 & 0.141 & RG & FRII & HERG & \checkmark & c & k;l & \\
  303.1 & 0.270 & RG & FRII-CSS & HERG &   & c-x & e & \\
  305.0 & 0.042 & RG & FRII & HERG &   & c-x & e & \\
  305.1 & 1.132 & RG & FRII-CSS & LERG &   & c &  & \\
  306.1 & 0.441 & RG & FRII & HERG & \checkmark & c & e & \\
  309.1 & 0.905 & QSO & CSS &  &   & c & e & \\
  310.0 & 0.054 & RG & FRI & LERG & \checkmark & c & igm & \\
  313.0 & 0.461 & RG & FRII & HERG & \checkmark & c & h;igm & \\
  314.1 & 0.120 & RG & FRI & LERG & \checkmark & c-x &  & \\
  315.0 & 0.108 & RG & FRI-XS & LERG & \checkmark & c & e & \\
  317.0 & 0.034 & RG & FRI & LERG & \checkmark & c-x & igm & igm\\
  318.0 & 1.574 & RG & FRII-CSS &  & \checkmark & c-x &  & \\
  318.1 & 0.045 & RG & FRI & LERG & \checkmark & c-x & igm & igm\\
  319.0 & 0.192 & RG & FRII & LERG & \checkmark & c-x &  & \\
  320.0 & 0.342 & RG & FRII & LERG & \checkmark & c & igm & \\
  321.0 & 0.096 & RG & FRII & HERG &   & c-x & h;l & \\
  322.0 & 1.681 & RG & FRII & HERG & \checkmark & x &  & igm\\
  323.0 & 0.679 & RG & FRII & HERG &   & c & e & \\
  323.1 & 0.264 & RG & FRII & HERG & \checkmark & c &  & \\
  324.0 & 1.206 & RG & FRII & HERG & \checkmark & c-x & e;h & \\
  325.0 & 1.135 & RG & FRII &  &   & c & h & \\
  326.0 & 0.090 & RG & FRII & LERG &   & c & l & \\
  326.1 & 1.825 & RG & FRII & HERG &   &   &  & \\
  327.0 & 0.105 & RG & FRII & HERG & \checkmark & c & h & \\
  327.1 & 0.462 & RG & FRI & HERG &   & c-x & k & \\
  330.0 & 0.55 & RG & FRII & HERG & \checkmark & c & h;l & \\
  332.0 & 0.151 & RG & FRII & HERG & \checkmark & c &  & \\
  334.0 & 0.555 & QSO & LDQ &  &   & c & k;l & \\
  336.0 & 0.927 & QSO &  &  &   & c &  & \\
  337.0 & 0.635 & RG & FRII & HERG & \checkmark & c-x & e & \\
  338.0 & 0.030 & RG & FRI & LERG & \checkmark & c-x & igm & igm\\
  340.0 & 0.775 & RG & FRII & HERG &   & c &  & \\
  341.0 & 0.448 & RG & FRII & HERG & \checkmark & c-x & e;k & igm\\
  343.0 & 0.988 & QSO & CSS &  &   & c &  & \\
  343.1 & 0.75 & RG & FRII-CSS &  &   & c &  & \\
  345.0 & 0.593 & QSO & CDQ &  & \checkmark & c-x & k & igm\\
  346.0 & 0.162 & RG & FRI & HERG & \checkmark & c & k & \\
  348.0 & 0.155 & RG & FRI & LERG & \checkmark & c-x & igm & igm\\
  349.0 & 0.205 & RG & FRII & HERG &   & c-x & h & \\
  351.0 & 0.372 & RG & FRII &  &   & c & h & \\
  352.0 & 0.807 & RG & FRII & HERG &   & c &  & \\
  353.0 & 0.030 & RG & FRII & LERG & \checkmark & c-x & k:l:igm & igm\\
  356.0 & 1.079 & RG & FRII & HERG &   & c & e & \\
  357.0 & 0.166 & RG & FRII & LERG & \checkmark & c &  & \\
  368.0 & 1.131 & RG & FRII & HERG &   & c &  & \\
  371.0 & 0.051 & BL & & - &   & c & k & \\
  379.1 & 0.256 & RG & FRII-XS & HERG &   & c &  & \\
  380.0 & 0.692 & QSO & CDQ &  &   & c & k & \\
  381.0 & 0.161 & RG & FRII & HERG &   & c &  & \\
  382.0 & 0.058 & RG & FRII & HERG &   & c-x &  & \\
  386.0 & 0.017 & RG & FRI & LERG & \checkmark & c-x &  & igm\\
  388.0 & 0.092 & RG & FRII & LERG & \checkmark & c & e & \\
  389.0 & ? & UNID & - & - &   & x &  & \\
  390.0 & ? & UNID & - & - &   &   &  & \\
  390.3 & 0.056 & RG & FRII & HERG &   & c-x & k;h & \\
  394.0 & ? & UNID & - & - &   &   &  & \\
  399.1 & ? & UNID & - & - &   &   &  & \\
  401.0 & 0.201 & RG & FRII & LERG & \checkmark & c & igm & \\
  402.0 & 0.026 & RG & FRI &  & \checkmark & c-x & k & \\
  403.0 & 0.059 & RG & FRII-XS & HERG &   & c & k;h & \\
  403.1 & 0.055 & RG & FRII & LERG & \checkmark & c &  & \\
  405.0 & 0.056 & RG & FRII &  & \checkmark & c-x & h;igm & igm\\
  409.0 & ? & UNID & - & - &   &   &  & \\
  410.0 & 0.249 & RG & FRII &  &   & c &  & \\
  411.0 & 0.467 & RG & FRII & HERG &   & c-x &  & \\
  415.2 & ? & UNID & - & - &   &   &  & \\
  418.0 & 1.686 & QSO &  &  &   &   &  & \\
  424.0 & 0.127 & RG & FRI & LERG & \checkmark & c & e & \\
  427.1 & 0.572 & RG & FRII & LERG & \checkmark & c & l;igm & \\
  428.0 & ? & UNID & - & - &   &   &  & \\
  430.0 & 0.056 & RG & FRII & LERG & \checkmark & c & e & \\
  431.0 & ? & UNID & - & - &   &   &  & \\
  432.0 & 1.785 & QSO &  &  &   & c-x &  & \\
  433.0 & 0.102 & RG & FRII-XS & HERG &   & c & l & \\
  434.0 & 0.322 & RG & FRII & LERG & \checkmark & c &  & \\
  435.0 & 0.471 & RG & FRII & HERG &   & c &  & \\
  436.0 & 0.215 & RG & FRII & HERG &   & c-x & e;h & \\
  437.0 & 1.48 & RG & FRII & HERG &   & c & e;h & \\
  438.0 & 0.29 & RG & FRII & HERG & \checkmark & c & igm & \\
  441.0 & 0.708 & RG & FRII & HERG &   & c &  & \\
  442.0 & 0.026 & RG & FRI & LERG & \checkmark & c & igm & \\
  445.0 & 0.056 & RG & FRII &  & \checkmark & c-x & h & \\
  449.0 & 0.017 & RG & FRI & LERG & \checkmark & c-x & igm & igm\\
  452.0 & 0.081 & RG & FRII & HERG & \checkmark & c-x & h;l & igm\\
  454.0 & 1.757 & QSO & CSS &  &   & x &  & \\
  454.1 & 1.841 & RG & FRII-CSS &  & \checkmark & &  & \\
  454.2 & ? & UNID & - & - &   &   &  & \\
  454.3 & 0.859 & QSO & CDQ &  &   & c-x & k & \\
  455.0 & 0.543 & QSO & CSS &  &   & c &  & \\
  456.0 & 0.233 & RG & FRII & HERG &   & c &  & \\
  458.0 & 0.289 & RG & FRII & HERG & \checkmark & c & h & \\
  459.0 & 0.220 & RG & FRII & HERG &   & c-x & l & \\
  460.0 & 0.268 & RG & FRII & LERG & \checkmark & c-x &  & \\
  465.0 & 0.030 & RG & FRI-WAT & LERG & \checkmark & c-x & k;igm & igm\\
  468.1 & ? & UNID & - & - &   &   &  & \\
  469.1 & 1.336 & RG & FRII & HERG &   & c-x &  & l\\
  470.0 & 1.653 & RG & FRII &  &   & c & h & \\
  
\noalign{\smallskip}
\end{longtable}
\endgroup
\end{center}

\footnotesize
\textbf{Notes.}\\
\tablenotemark{a} The 3CR name.\\
\tablenotemark{b} Redshift $z$. We also verified in the literature (e.g., NED/IPAC and/or SIMBAD databases) if updated $z$ values were reported after the release of the 3CR catalog.\\
\tablenotemark{c} The source classification: RG stands for radio galaxies, QSO for quasars; Sy for Seyfert galaxies and BL for BL Lac objects. We used the acronym UNID for sources that are still unidentified.\\
\tablenotemark{d} The radio morphological classification: FR\,I and FR\,II refer to the Fanaroff and Riley classification criterion \citep{fanaroff74} while LDQ and CDQ is sometimes adopted for lobe-dominated and core-dominated quasars; we also indicated if in the literature the source is classified as CSS or if the radio structure is X-shaped (XS), or if it is a narrow or wide angle tailed radio galaxy (NAT and WAT, respectively).\\
\tablenotemark{e} The optical spectroscopic designation: LERG, ``Low Excitation Radio Galaxy'', HERG, ``High Excitation Radio Galaxy''.\\
\tablenotemark{f} The ``cluster flag'' marks sources for which there is a known optical group/cluster reported in the literature and/or those for which there is an archival X-ray observation indicating the presence of hot gas.\\
\tablenotemark{g} The c flag indicates that at least one \chn\ observation is present in its archive while the x flag refers to the {\it XMM-Newton} archive.\\
\tablenotemark{h} In this column we report if the source has a radio component with an X-ray counterpart in a \chn\ observation. We used the following labels: $k$ = jet knot; $h$ = hotspot; $l$ = lobe; $e$ = extended X-ray emission around the nucleus and $igm$ = emission associated with the presence of hot gas.\\
\tablenotemark{i} The same as the previous column but for the {\it XMM-Newton} observations, we used only $l$ and $igm$.\\

\end{document}